\documentclass[structabstract]{aa}
\usepackage{graphicx}
\usepackage{txfonts}
\usepackage{subfigure}
\usepackage{natbib}

\def\HII{H\,{\sc ii} }

\begin{document}

   \title{On the kinematics of massive star forming regions: the case of
   IRAS\,17233--3606}

   \author{S. Leurini 
          \inst{1}
          \and
          C. Codella\inst{2} \and L. Zapata\inst{1,3} \and M.~T. Beltr{\'a}n\inst{2}  \and P. Schilke\inst{4} \and R. Cesaroni\inst{2}
          }

   \institute{Max-Planck-Institut f\"ur Radioastronomie,
              Auf dem H\"ugel 69, 53121 Bonn, Germany\\
              \email{sleurini@mpifr.de}
         \and INAF, Osservatorio Astrofisico di Arcetri, Largo E. Fermi 5, 50125 Firenze, Italy \and
Centro de Radioastronom\'ia y Astrof\'isica, Universidad Nacional Aut\'onoma de M\'exico, Morelia 58090, M\'exico 
\and Physikalisches Institut, Universit\"at zu K\"oln, Z\"ulpicher Str. 77, 50937 K\"oln, Germany}

   \date{\today}

  \abstract
   {Direct observations of accretion disks around high-mass young stellar objects would help to
discriminate between different models of formation of massive stars. However, given the complexity of massive star
forming regions, such studies are still limited in number. Additionally, there is still no general consensus on the molecular tracers to be used for such investigations.}
   {Because of its close distance and high luminosity, IRAS\,17233$-$3606 is a potential good laboratory to search for
traces of rotation in the inner gas around the protostar(s). 
Therefore, we selected the source for a detailed analysis of its molecular emission at 230~GHz with the SMA.}
   {We systematically investigated the velocity fields of transitions in the SMA spectra which are
not affected by overlap with other transitions, and  
 searched for coherent velocity gradients to compare them to the distribution of outflows in the region.  
Beside CO emission we also used high-angular H$_2$ images to trace 
the outflow motions driven by the IRAS\,17233$-$3606 cluster.}
   {We find linear velocity gradients in many transitions of the same 
molecular species and in several molecules. 
 We report the first detection of HNCO in molecular outflows from  massive YSOs.
We discuss the CH$_3$CN velocity gradient taking into account various
scenarios: rotation, presence of multiple unresolved
sources with different velocities, and outflow(s). Although other
interpretations cannot be ruled out, we propose that the CH$_3$CN
emission might be affected by the outflows of the region. Higher
angular observations are needed to discriminate between the different scenarios.}
{The present observations, with the possible association of CH$_3$CN with
outflows in a few thousands AU around the YSOs' cluster, (i) question  the choice of the tracer to probe rotating structures, and
(ii) show the importance  of the use of H$_2$ images for detailed studies of kinematics.}

   \maketitle

\section{Introduction}\label{intro}

Our understanding of the formation of massive stars has largely
improved in recent years. From a theoretical point of view, it has
been demonstrated, through 2--D and 3--D radiation hydrodynamic
simulations \citep{2009Sci...323..754K,2010ApJ...722.1556K}, that
radiation pressure is not a barrier to form stars with masses
$>$8~$M_{\rm \sun}$ and spectral type O and B and that they form
through disk accretion. In fact, the two main theoretical scenarios
proposed to explain the formation of massive stars are accretion-based
mechanisms: (1) the core accretion model
\citep{2002Natur.416...59M,2003ApJ...585..850M}, where a massive star
forms from a massive core fragmented from the natal molecular cloud,
and (2) the competitive accretion model \citep{2007prpl.conf..149B},
where a molecular cloud initially fragments into low-mass cores, which
form stars that compete to accrete mass from a common gas
reservoir. Both models predict the existence of protostellar accretion
disks around massive young stellar objects (YSOs), and therefore the
presence of molecular outflows, although the properties of both disks
and outflows could be different. In fact, the competitive accretion
model suggests that massive stars should always form in densely
clustered environments, and so the circumstellar disks could be
severely perturbed by interaction with stellar companions.

Over the last years, indirect evidence of the existence of accretion
disks have been provided by observations of phenomena such as
collimated outflows and jets with properties similar to those
originating from low-mass YSOs, thus implying a similar formation
mechanism for low- and high-mass stars.  However, direct observations
of accretion disks in high-mass star-forming regions are still limited
to few cases, mostly around YSOs with masses $\le 20~M_\odot$ and
luminosities $\le 10^4~L_\odot$ \citep[e.g., ][]{2010Natur.466..339K}.
The situation is less clear for more luminous objects
\citep{2007prpl.conf..197C},  although  recently \citet{2009ApJ...698.1422Z,2010ApJ...725.1091Z} 
suggested the presence of a Keplerian infalling ring around a central object
of at least 60~M$_\odot$.
This casts some doubts on the existence
of disks around O-type stars as no conclusive evidence for coherent, stable accretion disks around 
them as been presented 
so far. The clustered
mode of high-mass star formation and the typical large distances of
high-mass star forming regions challenge direct observation of
accretion disks.  In addition, it is still unclear which molecular
lines are best suited for such studies. They should have low optical
depths, and trace only the hot gas around the central protostar, and
not its envelope. In the past, molecules such as NH$_3$, CH$_3$CN,
HCOOCH$_3$, and C$^{34}$S have been proposed for this purpose
\citep[i.e.,
][]{1998ApJ...505L.151Z,1997A&A...325..725C,1999A&A...345..949C,18089_disk,
  2009ApJS..184..366B}.  However, there is still no general consensus
on the molecular tracer to use \citep[i.e., ][]{2005ApJ...628..800B}.

 IRAS\,17233$-$3606 (hereafter IRAS\,17233) is one of the best
laboratories to study the problem of massive star formation. \citet{2004A&A...426...97F} 
reported a luminosity of $1.7 \times 10^4~L_\odot$ obtained by
  integrating the spectral energy distribution and assuming $d=0.8$~kpc.
Distances between
700~pc and 2.2~kpc are reported in the literature \citep[i.e.,][]{2006A&A...460..721M,1989A&A...213..339F}.
However, the latter is based on maser emission lines. Using the  rotation curve determined by
\citet{1993A&A...275...67B} and velocities based on thermal lines \citep{1996A&AS..115...81B,1998AJ....116.1897M,2006A&A...460..721M,17233_irdc},
the kinematical distance of IRAS\,17233 is constrained to $\le 1$~kpc. The far distance ($\sim15$~kpc) is unlikely because of the source 
extremely bright  continuum
and line emission at practically all  observed wavelengths,
which would indicate exceedingly high luminosities if the source
were at the far distance. In the following, we assume a distance of 1~kpc for IRAS\,17233.

\begin{figure*}
\centering
\subfigure[][]{
\includegraphics[angle=-90,width=7cm]{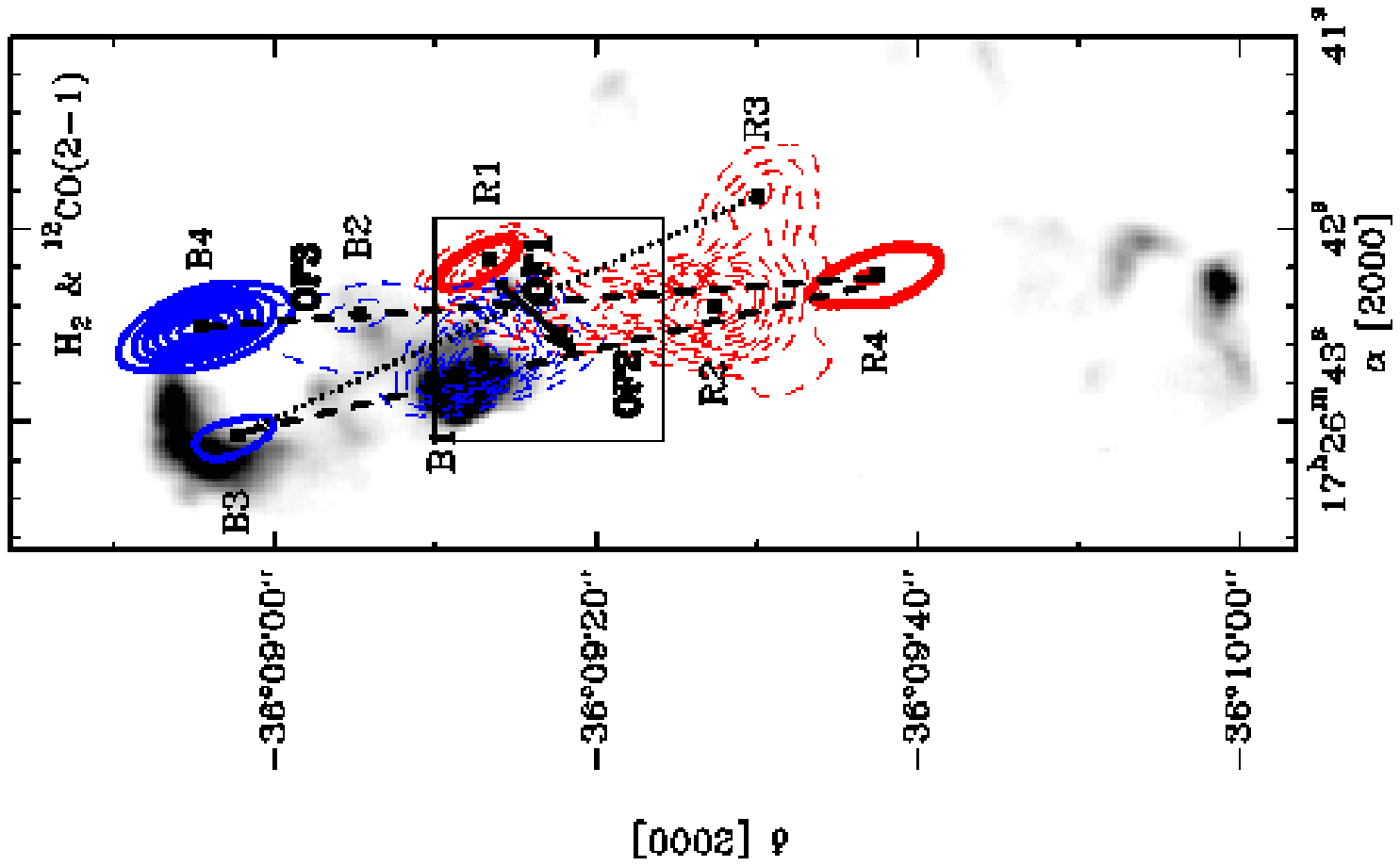}\label{jets}}
\subfigure[][]{
\includegraphics[angle=-90,width=8cm]{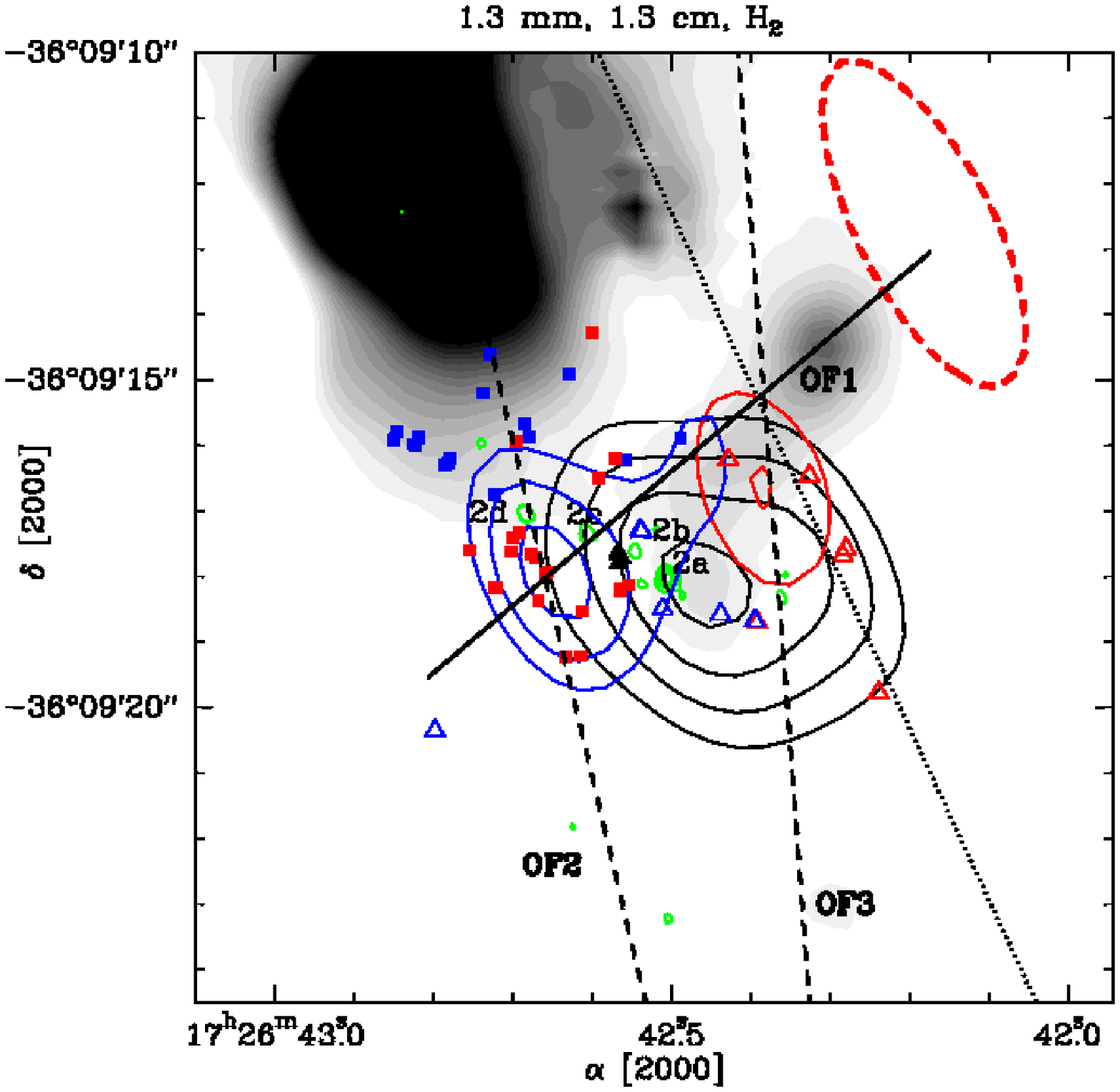}\label{zoom}}
\caption{{\it Left panel:} Integrated emission of the blue- and red-shifted wings in
the CO(2--1) line overlaid on the H$_2$ emission at 2.12$\mu$m (grey scale). 
 The solid contours show the EHV blue-
(v=[-200,-130]~km~s$^{-1}$) and red-shifted emission
(v=[90,120]~km~s$^{-1}$); the dashed contours mark the HV blue-
(v=[-130,-25]~km~s$^{-1}$) and red-shifted emission
(v=[16,50]~km~s$^{-1}$).      
The dashed lines and the solid arrow  outline the  possible 
molecular outflows identified in CO (OF1, OF2, OF3, 
see discussion in Sect.~\ref{i17233}); 
the dotted line marks the alternative direction of the OF2 outflow. 
The central box outlines the region mapped in the right panel. The squares mark the positions of 
the red- and blue-shifted CO(2--1) peaks.
{\it Right panel:} 
Continuum emission at 1.3~mm from the maser zone in
  IRAS\,17233 (solid black contours from $50\%$ of peak emission in step of $15\%$, see 
Sect.~\ref{sec_con}).  
Overlaid on the map
  are the positions of the OH \citep[blue and red 
    squares,][]{2005ApJS..160..220F}, CH$_3$OH \citep[black
    triangles,][]{1998MNRAS.301..640W} and water \citep[blue
    and red open triangles,][]{2008AJ....136.1455Z} masers. The green
  contour levels show the continuum emission at 1.3~cm \citep[from
    1.4~mJy~beam$^{-1}$, $3\sigma$, in step of
    1.4~mJy~beam$^{-1}$;][]{2008AJ....136.1455Z}, while the grey scale
  map is the H$_2$ emission. The solid blue and red contours represent the SO($5_6-4_5$) blue- and red-shifted emission 
(blue: from 80 to 104~Jy~km~s$^{-1}$~beam$^{-1}$ in step of 10~Jy~km~s$^{-1}$~beam$^{-1}$; red: 10 and 12~Jy~kms$^{-1}$~beam$^{-1}$). The dashed red contour shows the EHV red-shifted CO($2-1$) emission.
 OF1, OF2 and OF3 are marked as in left panel.
\label{overview}}
\end{figure*}

 Thanks to
combined high angular resolution observations of different tracers,
\citet{2009A&A...507.1443L} found evidence for multiple outflows
driven by different hyper-compact \HII (HC\HII) regions resolved at milliarcsec-scale with
VLA by \cite{2008AJ....136.1455Z} (see Fig.~\ref{overview}). Their
parameters are typical of massive YSOs (early B-type ZAMS stars) and in agreement with the
measured bolometric luminosity of the source. 
\citet{2009ApJS..184..366B} found evidence for a
large-scale rotating structure detected in high excitation lines of
NH$_3$ along the ESE--WNW direction. On the other hand, the high
spatial-resolution maps in CO and H$_2$ by \citet{2009A&A...507.1443L} indicated a
compact bipolar outflow located along the NW-SE, thus questioning the
interpretation of the NH$_3$ gradient as a rotating structure.

We present here observations of the massive star forming region
IRAS\,17233 in several molecular tracers around 230~GHz taken with the
Submillimeter Array. The idea of this study is to analyse the spatial
distribution and velocity field of different lines, compare them to
the distribution of outflows in the region, and verify whether we find
evidence for rotating structures perpendicular to the outflows or
not. Given the large bandwidth of SMA spectra ($\sim4$~GHz) and the
richness of the molecular spectrum of the source, such observations
allow us to study the distribution of several lines, and not only of
the ``best'' disk tracer candidates.

\section{Molecular outflows from the YSOs in IRAS\,17233$-$3606}\label{i17233}

In order to investigate the presence of possible accretion disks around the protostars
in the inner region of IRAS\,17233, a detailed 
knowledge of the molecular outflows in the area is needed.
In \citet{2009A&A...507.1443L} we analysed the emission of CO isotopologues observed simultaneously to the
data presented in this paper (see Sect.~\ref{obs}) and compared their distribution to  high-spatial resolution H$_2$ observations.
The main results of that study
can be summarised in the following points (see Fig. 1):

\begin{enumerate}

\item
The data reveal a clumpy extended structure with
well separated blue- and red-shifted emission, and an overall
structure roughly aligned along the N--S direction.  
The outflow is associated with
extremely high velocities, up to $\sim$ --200 and +120 km s$^{-1}$
with respect to the ambient LSR velocity.

\item
Using high angular resolution maps of H$_2$, cm-continuum, and maser emission,
multiple outflows can
be distinguished in the region. 
One (called here OF1) is compact ($\sim$ 5$\arcsec$-10$\arcsec$) and located
along the NW-SE direction,  as detected in CO and SO: it shows extremely 
high-velocity red-shifted emission and it is associated with a jet traced by H$_2$ emission
as well as with red and blue H$_2$O maser spots.
The extended emission along the N--S direction is  probably due to multiple outflows. 
The picture derived by connecting the position of the CO
high velocity clumps suggests two possible main axes, here labelled as OF2 and OF3
and drawn in Fig.~\ref{overview} by dashed lines. 
Interestingly, OF2 is associated  
with a counterpart on smaller scales ($\sim5''$, corresponding to $\sim$5000~AU at $D=1$~kpc) 
traced by red- and blue-shifted OH masers (Fig.~\ref{zoom}). 
In addition, another possible outflow direction is indicated in Fig.~\ref{overview} by the
dotted line.

\item
The powering sources of the flows remain unidentified, as they 
are centred on a
region where 4 HC\HII
regions (VLA~2a, b, c and d, see Fig.~\ref{zoom} and ~\ref{cont}) 
have been observed on a sub-arcsec scale.

\end{enumerate}

\section{Observations}\label{obs}

We observed IRAS\,17233 with the SMA interferometer on April
10, 2007 in the compact configuration with seven
antennas.  The receivers operated in a double-sideband mode with an IF
band of 4-6 GHz so that the upper and lower sideband (USB and LSB, respectively) were separated by
10 GHz. The central frequencies of the upper and lower sideband were
220.4 and 230.4~GHz, respectively.

The observations presented in this study were performed simultaneously to those of CO discussed by \citet{2009A&A...507.1443L}. 
For details on the weather conditions, calibration, and reduction of the dataset we refer to their Sect. 2. 
 We checked the absolute positional uncertainty  on the quasar 3C273, our phase calibrator, and found it to be $<0\farcs8$
for the right ascension and $<0\farcs5$ for the declination.

The  resulting synthesised beams, the  velocity resolution used in the analysis and the typical r.m.s of the maps are given in Table~\ref{obs-para}. 
The centre of the maps is at $\alpha_{2000}$=17$^h$26$^m$41$^s$.757,
$\delta_{2000}$=$-$36$^{\circ}$09$'$0$\farcs$50. 
The conversion factor from flux density to brightness temperature in the synthesised beam is $\sim$2.25 K~(Jy/beam)$^{-1}$ for the lower side band, and $\sim$2.53 K~(Jy/beam)$^{-1}$ for the upper side band data. 
\begin{table}
\centering
\caption{Observational parameters.\label{obs-para}}
\begin{tabular}{rccccl}
\hline
\multicolumn{1}{c}{} &\multicolumn{1}{c}{HPBW}&\multicolumn{1}{c}{P.A.}
&\multicolumn{1}{c}{$\Delta \rm{v}$}&\multicolumn{1}{c}{r.m.s.}\\
&\multicolumn{1}{c}{($\arcsec$)}&\multicolumn{1}{c}{($^\circ$)}&\multicolumn{1}{c}{(${\rm km~s}^{-1}$)}
&\multicolumn{1}{c}{(${\rm Jy~beam}^{-1}$)}\\
\hline
cont.&$ 4\farcs9\times1\farcs8$&$\sim 29^\circ$&&0.04\\
LSB&$5\farcs4\times1\farcs9$&$\sim 29^\circ$&0.5&0.1\\
USB&$4\farcs9\times1\farcs8$&$\sim 29^\circ$&0.5&0.1\\
\hline
\end{tabular}
\end{table}

\section{Observational results}

\subsection{Continuum emission}\label{sec_con}

\begin{figure}
\centering
\includegraphics[angle=-90,width=9cm]{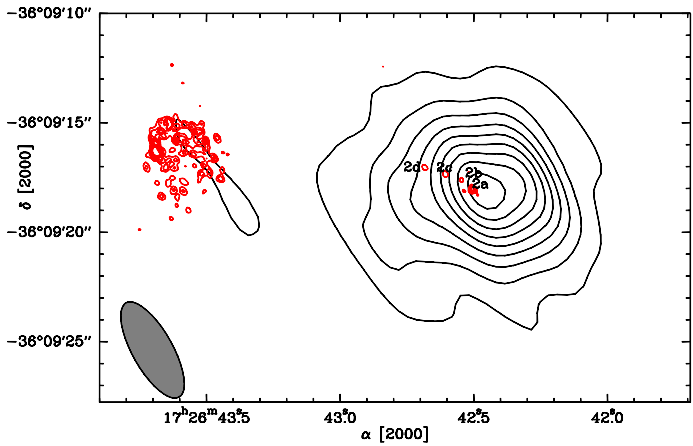}
\caption{
Continuum emission at 1.3~mm (black contours) and 1.3~cm (red contours) of IRAS\,17233. Contours are 
from   2~mJy~beam$^{-1}$ ($5\sigma$) in step of    2~mJy~beam$^{-1}$ (1.3~cm emission), 
and from 0.12~Jy~beam$^{-1}$  ($3\sigma$) in steps of 0.24~Jy~beam$^{-1}$ (1.3~mm emission). The four HC\HII VLA~2a, b, c and d identified by
\cite{2008AJ....136.1455Z} are also labelled.\label{cont}}
\end{figure}

The  continuum  emission   of the region at  1.3~mm     is  shown  in
Fig.~\ref{cont}.  This was derived using only the upper side band data, which have a lower number 
of lines than  the lower side band data and therefore line subtraction is less  problematic.
The peak  of  the  emission is  found  close to  the
cluster  of  four  compact   sources  detected  at  cm-wavelengths  by
\citet{2008AJ....136.1455Z}.  This also  corresponds to  the so-called
``maser  zone'', where  strong  H$_2$O, CH$_3$OH,  and  OH masers  are
detected \citep[e.g.,][see Fig.~\ref{zoom}]{1980IAUC.3509....2C,1982ApJ...259..657F,1991ApJ...380L..75M}. 
Weak
($\sim 3\sigma$) emission  is also detected from the  \HII region on the east of the maser zone, 
already    mapped    at    cm-wavelengths   by    several   authors
\citep[e.g.,][]{1993AJ....105.1495H,1998MNRAS.301..640W}. 

We performed a 2-D Gaussian fit of the 1.3~mm continuum emission, and
determined its peak to be at $\alpha_{\rm J2000}=17^h26^m42.455^s$,
$\delta_{\rm J2000}=-36^\circ09'18.047\arcsec$ with  positional errors of $0\farcs08$.
The peak flux ($F_\nu$) is $\sim
2.09\pm0.07$~Jy beam$^{-1}$, the integrated flux ($S_\nu$) 7.20 Jy.  
However, given the richness of the molecular spectrum
of the region at this frequency, the continuum emission may have been
overestimated because of line contamination. 
The inferred source size is $6\farcs1\pm 0\farcs2\times 5\farcs1\pm 0\farcs2$ 
with a P.A. of $64^\circ.2 \pm 7^\circ.5$, corresponding to a deconvolved size of $5\farcs3\times2\farcs7$, and a  P.A. of $-77^\circ.2$ (see Table~\ref{fit}).

The peak of the mm continuum emission does not correspond to any of the compact sources
detected by \citet{2008AJ....136.1455Z} at 1.3~cm, but it is shifted to the west.
VLA~2a, b, c and d are all within $\sim 2\arcsec$ 
from the peak of the mm continuum emission. However, the displacement between the mm continuum peak and the strongest 
cm source VLA~2a is only $\sim$ $(0\farcs6,0\farcs0)$, 
while the offset from  the CH$_3$OH maser spots \citep[whose uncertainty in
 the absolute position is $\sim 1''$,][]{1998MNRAS.301..640W} is $\sim(1\farcs4,0\farcs3)$. 
Given the uncertainty in the
absolute position of our SMA data ($\la0\farcs8$), this may not be significative.

We can calculate beam-averaged gas column density and mass using the
equations given by \citet{2002ApJ...566..945B} corrected in the
erratum by \citet{2005ApJ...633..535B}. We assume the authors' default
values for the grain size (0.1 $\mu$m), grain mass density (3
g~cm$^{-3}$),  gas-to-dust ratio (100) and  a typical dust temperatures of 200~K for hot cores. 
For the grain emissivity, we used $Q_\nu=7.5\times10^{-4}(125\mu m/\lambda)^\beta, \beta=1.6$, as given by \citet{2005ApJ...633..535B}.
 We derived a beam-averaged  H$_2$ column density of 
1.0$\times 10^{24}$~cm$^{-2}$, and a mass of 12 $M_\odot$. 
However, the mass and column density
derived in this fashion are affected by the uncertainties in the peak
 and integrated flux determinations due to line
contamination.  Based on unpublished SMA observations at 217~GHz in
the same configuration of the array as for the current data (Leurini
et al. in prep.), we derived a second independent measurement of
$S_\nu$ and $F_\nu$ at a wavelength close to that presented here. The peak flux at 217~GHz is estimated to be $2.10\pm 0.04$ Jy beam$^{-1}$  
(on a beam of $3\farcs31 \times 2\farcs54$), while the integrated intensity 
of the source is 6.69 Jy. These values are very close to those derived at 230~GHz, with an
uncertainty of less than 10\% in the peak flux (but on a different beam) and on the integrated intensity. Therefore, the
uncertainties in the estimates of the mass and of the column density are dominated by the uncertainty in the dust temperature.

\subsection{Line emission}\label{lte}

\begin{figure*}
\centering
\subfigure[]{\includegraphics[angle=-90,width=17cm, bb = 200 15 552 690,clip]{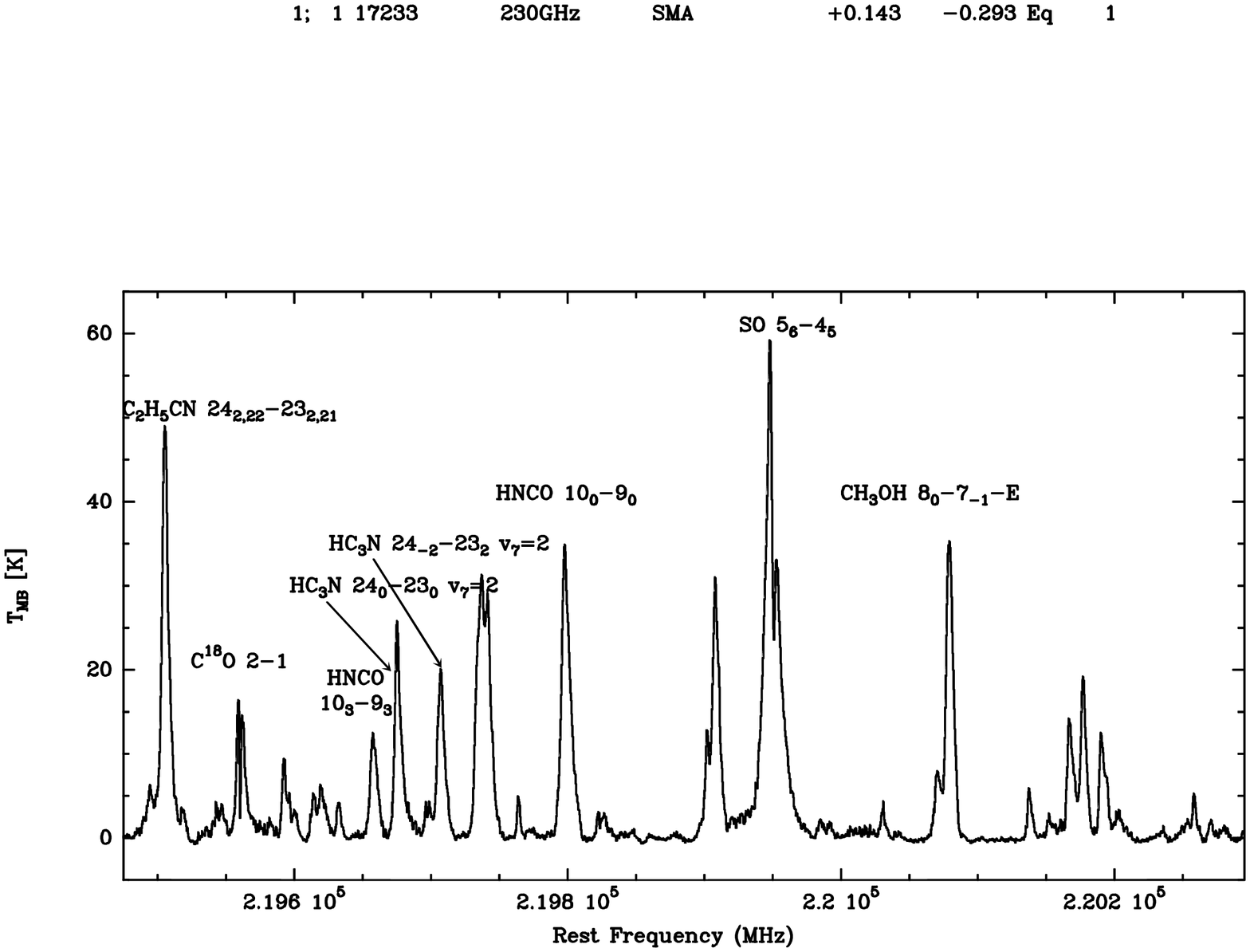}}
\subfigure[]{\includegraphics[angle=-90,width=17cm, bb = 200 15 552 690,clip]{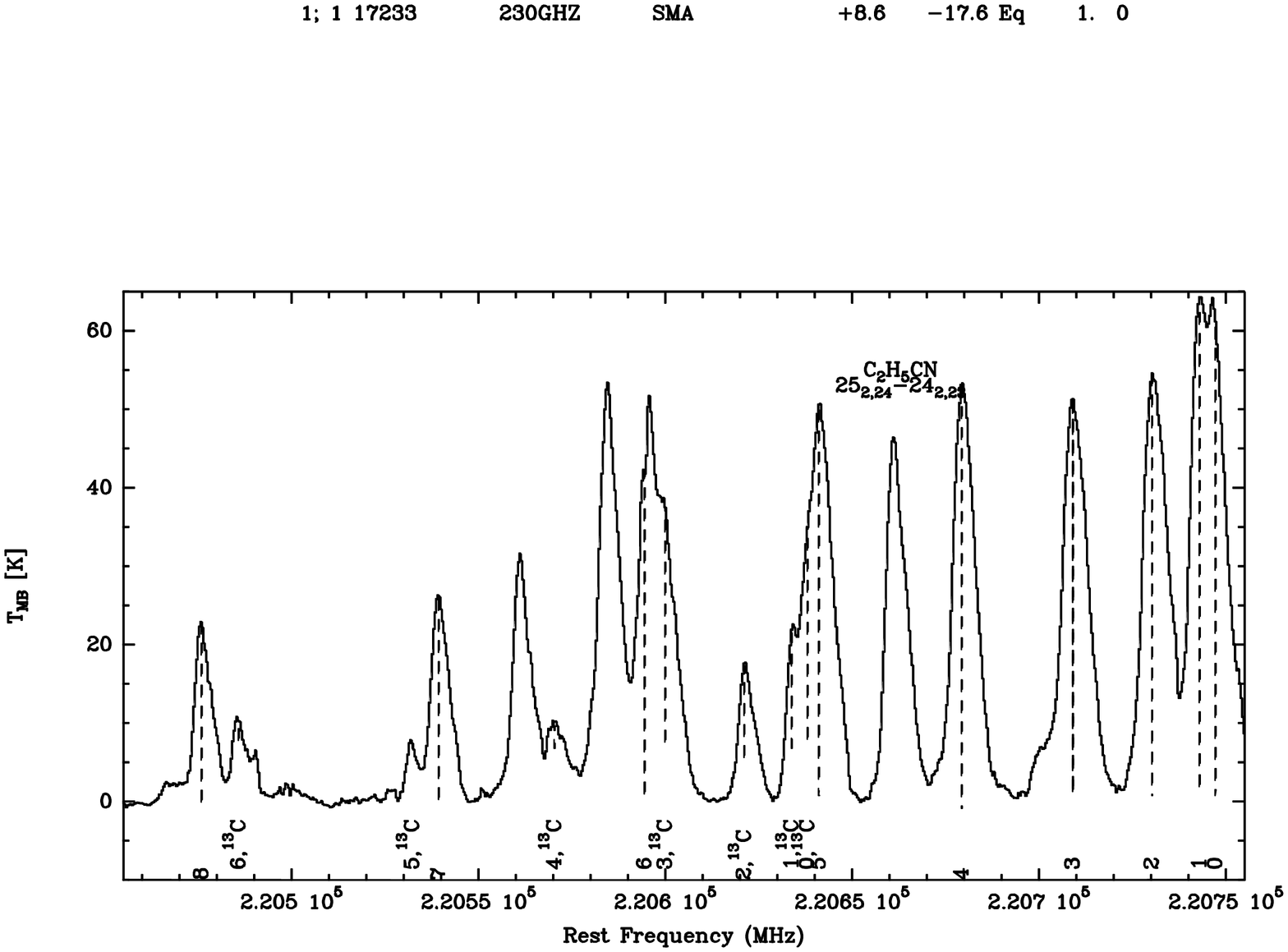}}
\caption{Lower side band  continuum-subtracted representative spectra of IRAS\,17233 at the peak of the 1.3~mm continuum emission. All transitions used 
to derive the zero-th and first moment maps of Figs.~\ref{ch3ohmom0}-\ref{hncomom1} are labelled. For reference, we also mark the C$^{18}$O ($2-1$) and SO$(5_6-5_5)$ lines 
\citep[discussed in][]{2009A&A...507.1443L}, and the ladder of the 
CH$_3$CN (labelled with the corresponding $K$ quantic number) and CH$_3^{13}$CN ($12-11$) lines (labelled with the corresponding $K$ quantic number followed by $^{13}$C).\label{spectra-lsb}}
\end{figure*}

The molecular spectrum associated with the peak of the 1.3~mm continuum emission is very rich in 
lines, 
and typical of hot cores in massive YSOs. The LSB spectrum of the source (Fig.~\ref{spectra-lsb}) is dominated by 
the CH$_3$CN ($12_K-11_K$) band: transitions up to $K=9$ are detected, 
together with lines from vibrationally excited levels (up to $K=7$), 
and from the CH$_3^{13}$CN isotopologue (up to $K=6$). This 
points to high excitation conditions since the lower level energies of these lines range up to 
840~K. 
The detection frequency of lines in the USB spectrum  is less than for the LSB band, and the band 
is dominated by the CO ($2-1$) transition.

Given the complexity of the region and the richness of the spectrum,
line identification in molecular spectra of hot cores is 
challenging. In the analysis of the spectrum of IRAS\,17233, we made 
use of the XCLASS program\footnote{http://www.astro.uni-koeln.de/projects/schilke/XCLASS}.
The method used in the program is 
 described by \citet{2005ApJS..156..127C} (and references 
therein), and  it is based on a simultaneous fit of all lines in a spectrum 
under  the LTE
approximation. The model takes into account the beam dilution, the
line opacity, and the line blending, and produces a synthetic spectrum of the source for given 
physical parameters ($T_{\rm{kin}}$, $N_{\rm{species}}$, source size). The molecular spectroscopic
parameters are taken from the CDMS catalogue
\citep{2001A&A...370L..49M,2005JMoSt.742..215M} and from the molecular
spectroscopic database of the Jet Propulsion Laboratory \citep[JPL,
  see][]{pickett_JMolSpectrosc_60_883_1998}.  

Although this approach
delivers physical quantities such us rotational temperature, column
density and source size for each of the molecular species included in
the model, we believe that the complexity of the region is too high to
infer meaningful results with such a simple approach. The analysis of the hot core chemistry of the source
is beyond the scope of this paper and will be treated in a dedicated study. 
Nevertheless, a LTE synthetic spectrum,
obtained with plausible input values for the physics of the gas of the
inner region around the protostars, is fundamental for a correct
identification of spectral features, since it predicts intensities for
any transition of a given molecular species at all known frequencies.
Based on the line profiles, and on the distribution of several
transitions which show extended emission (see Sect.~\ref{mom0}), we
assumed two components inside the beam, the first  with a compact
($\le 2''$) source with a temperature of $\sim$200~K and a line width
of 4~km~s$^{-1}$, and a second more extended ($\sim 10''$) source with
lower temperature ($50-70$~K) and broader line width ($\sim
10$~km~s$^{-1}$).   Table~\ref{id} lists all transitions identified in
the LSB and USB spectra of IRAS\,17233 towards the peak of the mm
continuum emission, which are analysed in this paper, and their line parameters obtained by using
 Gaussian profiles.

\begin{table*}
\centering
\caption{Spectral features detected in the IRAS\,17233 spectrum  and analysed in the present paper}\label{id}
\begin{tabular}{lcrrrr}
\hline
\hline   
\multicolumn{1}{c}{transition}&\multicolumn{1}{c}{$\nu$} &\multicolumn{1}{c}{$E_u$}&\multicolumn{1}{c}{$\rm{v}_{\rm{LSR}}$}&\multicolumn{1}{c}{$\Delta \rm{v}$}&\multicolumn{1}{c}{$\int{T\delta \rm{v}}$}\\
&\multicolumn{1}{c}{[MHz]} &[K]&[km s$^{-1}$]&[km s$^{-1}$]&[K km s$^{-1}$]\\
\hline

C$_2$H$_5$CN$(24_{2,22}-23_{2,21})$&219505.59&106&$-3.40\pm0.11$&$7.96\pm0.29$&$386.6\pm11.0$\\
HNCO $(10_3-9_3)$&219656.77&433&$-5.23\pm0.04$&$8.11\pm0.10$&$100.0\pm1.0$\\
HC$_3$N $(24_0-23_0)~v_7=2$&219675.11&773&$-4.29\pm0.33$&$7.45\pm0.89$&$178.8\pm16.5$\\
HC$_3$N $(24_{-2}-23_2)~v_7=2$&219707.35&778&$-3.28\pm0.14$&$8.45\pm0.40$&$163.9\pm6.0$\\
HNCO $(10_0-9_0)$&219798.27&58&$-3.94\pm0.01$&$10.16\pm0.03$&$342.0\pm1.0$\\
CH$_3$OH-$E$ $(8_0-7_{-1})$&220078.49&97&$-4.56\pm0.12$&$8.36\pm0.32$&$304.8\pm9.2$\\
CH$_3$CN $(12_8-11_8)$&220475.92&525&$-4.02\pm0.05$&$7.44\pm0.12$&$169.8.8\pm2.2$\\
CH$_3$CN $(12_7-11_7)$&220539.41&419&$-3.85\pm0.43$&$7.79\pm1.07$&$210.8\pm24.2$\\
CH$_3^{13}$CN $(12_2-11_2)$&220621.14&97&$-4.33\pm0.75$&$6.98\pm1.84$&$123.6\pm27.0$\\
C$_2$H$_5$CN $(25_{2,24}-24_{2,23})$&220660.92&121&$-4.64\pm0.58$&$8.69\pm1.39$&$407.0\pm54.3$\\
CH$_3$CN $(12_4-11_4)$&220679.32&183&$-4.29\pm0.46$&$9.30\pm1.14$&$515.5\pm52.5$\\
CH$_3$CN $(12_2-11_2)~v_8=1$&221367.67&596&$-3.63\pm0.37$&$8.14\pm0.97$&$178.5\pm17.3$\\
CH$_3$OH-$E$ $(8_{-1}-7_0)$&229758.76&89&$-3.22\pm0.02$&$7.94\pm0.04$&$484.0\pm2.1$\\
O$^{13}$CS $(19-18)$&230317.53&99&$-2.50\pm0.04$&$6.36\pm0.11$&$118.9\pm1.7$\\
OCS $(19-18)$&231060.99&100&$-3.04\pm0.02$&$10.51\pm0.04$&$574.2\pm1.7$\\
CH$_3$OH-$A$ $(10_2-9_3)$&231281.10&165&$-3.72\pm0.03$&$7.31\pm0.08$&$165.6\pm1.3$\\
\hline

\end{tabular}
\end{table*}

\begin{table*}
\centering
\caption{Result of the 2-D Gaussian fit of the integrated intensity distributions.}\label{fit}
\begin{tabular}{lccccc}
\hline
\hline   
\multicolumn{1}{c}{transition}&\multicolumn{1}{c}{$\Delta \alpha^{\mathrm{a}}$} &\multicolumn{1}{c}{$\Delta \delta^{\mathrm{a}}$}&\multicolumn{1}{c}{FWHM observed size$^\mathrm{b}$}&\multicolumn{1}{c}{FWHM deconvolved size}\\
\hline
cont.&$8\farcs455$&$-17\farcs551$&$6\farcs1\times5\farcs1$&$5\farcs3\times2\farcs7$\\

C$_2$H$_5$CN$(24_{2,22}-23_{2,21})$&$8\farcs930$&$-17\farcs538$&$5\farcs9\times3\farcs2$&$2\farcs6\times2\farcs2$\\
HNCO $(10_3-9_3$)&$8\farcs987$&$-17\farcs140$&$5\farcs0\times 1\farcs9$\\
HC$_3$N $(24_0-23_0)~v_7=2$&$8\farcs941$&$-17\farcs421$&$5\farcs5\times 2\farcs3$&$1\farcs5\times 0\farcs4$\\
HC$_3$N $(24_{-2}-23_2)~v_7=2$&$8\farcs832$&$-17\farcs475$&$5\farcs4 \times2\farcs2$\\
HNCO $(10_0-9_0$)&$8\farcs830$&$-17\farcs399$&$5\farcs9\times 2\farcs9$&$2\farcs5\times 2\farcs1$\\
CH$_3$OH-$E$ $(8_0-7_{-1}$)&$9\farcs091$&$-17\farcs108$&$5\farcs9\times 3\farcs9$&$3\farcs6\times 2\farcs2$\\
CH$_3$CN $(12_8-11_8$)&$8\farcs885$&$-17\farcs245$&$5\farcs5\times 2\farcs6$&$2\farcs0\times 0\farcs4$\\
CH$_3$CN $(12_7-11_7$)&$9\farcs010$&$-17\farcs305$&$5\farcs5\times 2\farcs9$&$2\farcs2\times 1\farcs0 $\\
CH$_3^{13}$CN $(12_2-11_2$)&$8\farcs900$&$-17\farcs400$&$5\farcs4\times 3\farcs0$\\
C$_2$H$_5$CN $(25_{2,24}-24_{2,23}$)&$9\farcs076$&$-17\farcs459$&$5\farcs8\times 3\farcs1$&$2\farcs6  \times  2\farcs0$\\
CH$_3$CN $(12_4-11_4$)&$9\farcs113$&$-17\farcs386$&$5\farcs9 \times3\farcs8$&$3\farcs3 \times2\farcs2$\\
CH$_3$CN $(12_2-11_2)~v_8=1$&$8\farcs975$&$-17\farcs375$&$5\farcs3\times 2\farcs4$\\
CH$_3$OH-$E$ $(8_{-1}-7_0$)&$9\farcs247$&$-16\farcs994$&$5\farcs8\times 4\farcs7$&$4\farcs4\times 2\farcs9$\\
O$^{13}$CS $(19-18$)&$8\farcs850$&$-17\farcs559$&$5\farcs2\times 3\farcs6$&$3\farcs1\times 1\farcs7$\\
OCS $(19-18$)&$9\farcs216$&$-17\farcs424$&$5\farcs7\times 4\farcs4$&$4\farcs0\times 2\farcs9$\\
CH$_3$OH-$A$ $(10_2-9_3$)&$9\farcs173$&$-17\farcs082$&$5\farcs3\times 3\farcs8$&$3\farcs3\times 2\farcs0$\\
\hline
\end{tabular}
\begin{list}{}{}
\item[$^{\mathrm{a}}$] offset from the centre of the maps at
$\alpha_{2000}$=17$^h$26$^m$41$^s$.757,
$\delta_{2000}$=$-$36$^{\circ}$09$'$0$\farcs$50 (see Sect. 3).
\item[$^{\mathrm{b}}$] typical errors on the fit are of the order of 0\farcs2.
\end{list}
\end{table*}

\section{Data analysis}\label{res}
The frequency setup of our observations covers the CH$_3$CN
($12_K-11_K$) band, used in the past to trace disks and/or toroids
around massive YSOs, as well as species which have been detected in
molecular outflows from low- and high-mass protostars (i.e.,
H$_2^{13}$CO, CH$_3$OH, OCS). In addition, several high energy
transitions are detected, which are potentially good candidates to
search for tracers of rotation.

To study the velocity field traced by different molecular features, we
derived maps of the zero-th and first moments of different
transitions. We limited our analysis to lines which are not affected
by overlap with other transitions.  This was verified with the help of
our LTE model (see \S~\ref{lte}).  
However, since several
spectral features remain unidentified,  
it is possible that lines used to derive the
distribution of the zero-th and first moments are  affected by
emission from species not included in our model. 

Table~\ref{id} lists all transitions identified in the LSB and USB
spectra of IRAS\,17233 towards the peak of the mm continuum emission
and used to derive the zero-th and first moment maps. Part of the LSB spectrum towards the peak of the continuum is  
shown
in Fig.~\ref{spectra-lsb} as example of the complexity of the source.
Figures~\ref{ch3ohmom0}-\ref{ch3cnmom1} and \ref{ocsmom1}-\ref{hncomom1} show the derived distributions of
the zero-th order and first moments computed over  a region where the
signal-to-noise ratio in the line data cubes is larger than  $5\sigma$ (corresponding to
0.5~Jy~beam$^{-1}$).

\subsection{Spatial distribution}\label{mom0}
\begin{figure}
\centering
\includegraphics[angle=-90,width=8cm]{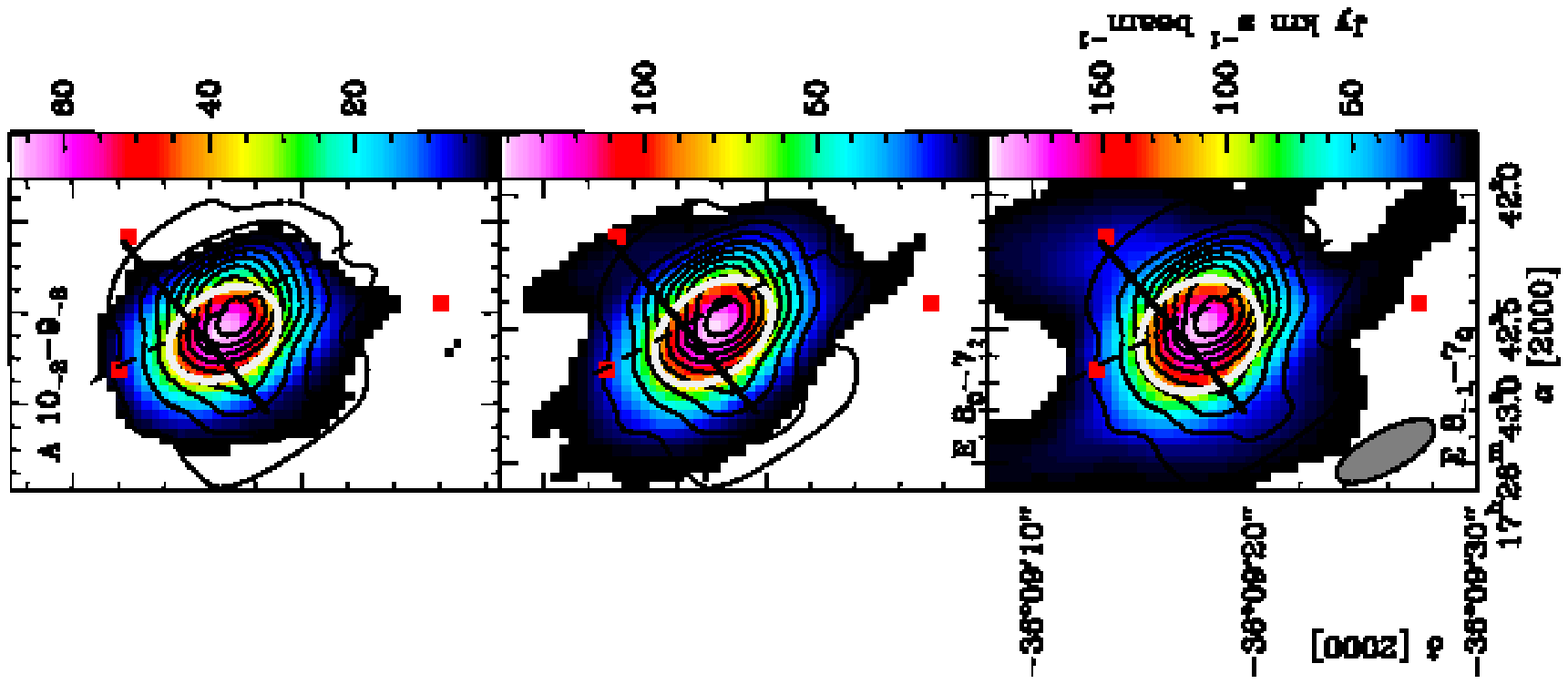}
\caption{Integrated emission of selected transitions of CH$_3$OH in the region above the  
 5$\sigma$ detection in the line data cubes ($\sim$ 0.5 Jy~beam$^{-1}$).
The black solid contours show the 1.3~mm continuum emission as in Fig.~\ref{cont}. 
The solid line shows the axis of OF1, while the dashed line has been chosen,
for sake of clarity, to represent
the three possible directions of the N--S outflows drawn in Fig.~\ref{overview}. 
The red squares mark  the R1, R2 and B1 peaks of the CO ($2-1$) emission detected towards OF1 and OF2--OF3  (see Fig.~\ref{jets}).
The white ellipse represents the FWHM 2-D Gaussian fit to the data. 
\label{ch3ohmom0}}
\end{figure}

\begin{figure}
\centering
\includegraphics[width=8cm,angle=-90]{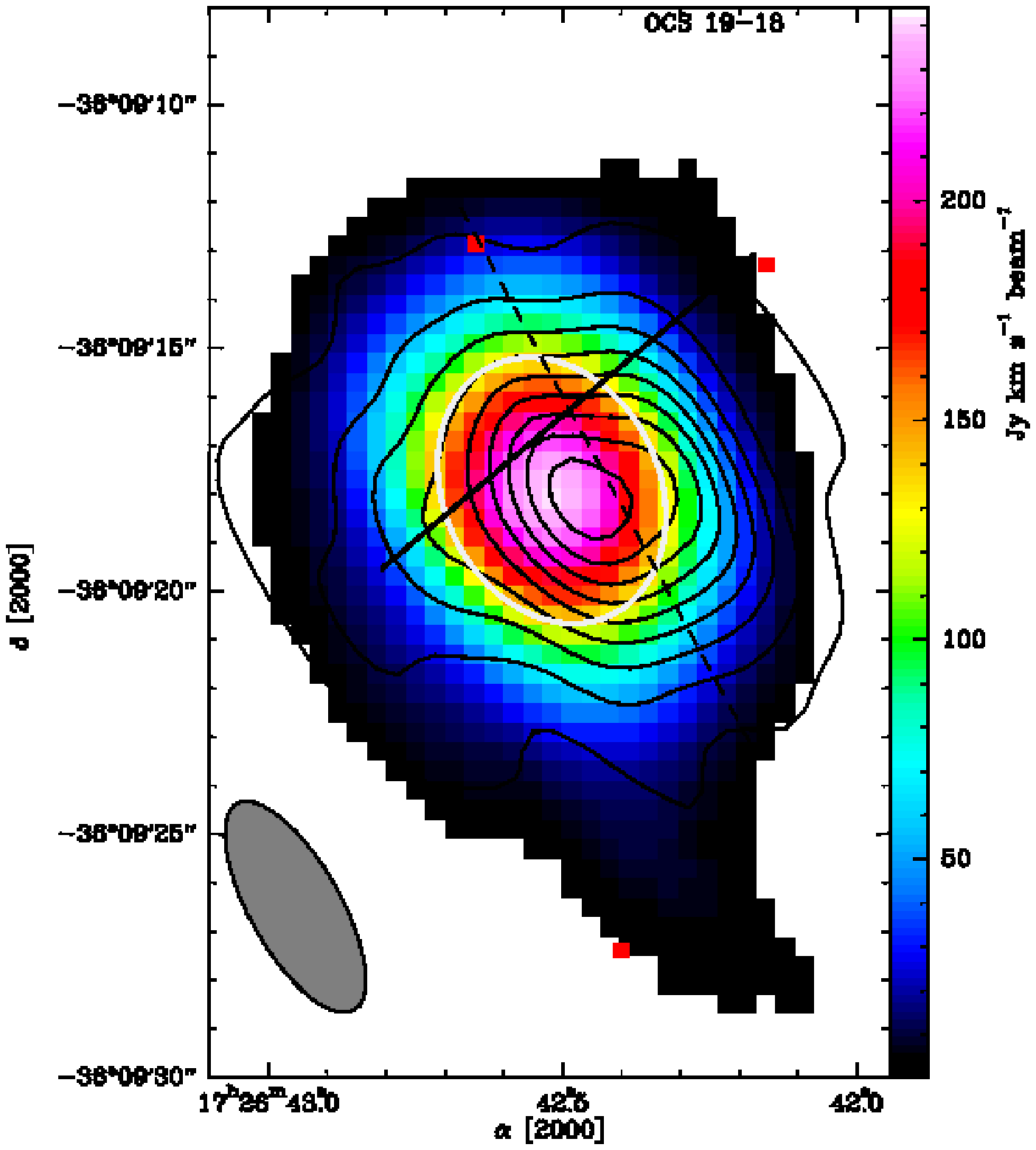}
\caption{Integrated emission of the OCS ($19-18$) line  in the region above the  
5$\sigma$ detection in the line data cubes ($\sim$ 0.5 Jy~beam$^{-1}$).
The black solid contours show the 1.3~mm continuum emission as in Fig.~\ref{cont}. 
Symbols are as in Fig.~\ref{ch3ohmom0}.
\label{ocsmom0}}
\end{figure}

\begin{figure}
\centering
\includegraphics[angle=-90,width=8cm]{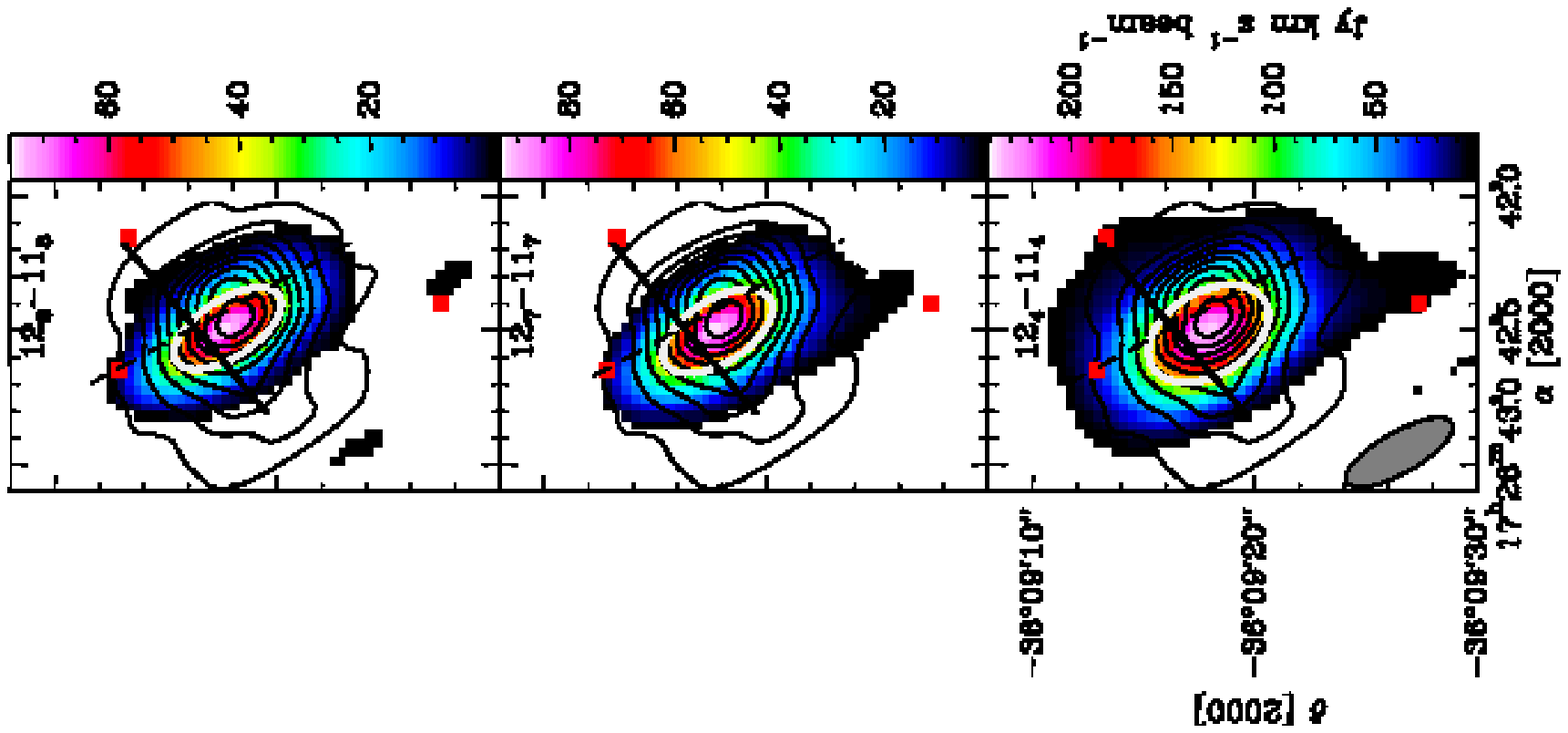}
\caption{Integrated emission of selected transitions of CH$_3$CN 
 in the region above the  
5$\sigma$ detection in the line data cubes ($\sim$ 0.5 Jy~beam$^{-1}$).
The black solid contours show the 1.3~mm continuum emission. 
Symbols are as in Fig.~\ref{ch3ohmom0}. \label{ch3cnmom0}}
\end{figure}

\begin{figure}
\centering
\includegraphics[angle=-90,width=8cm]{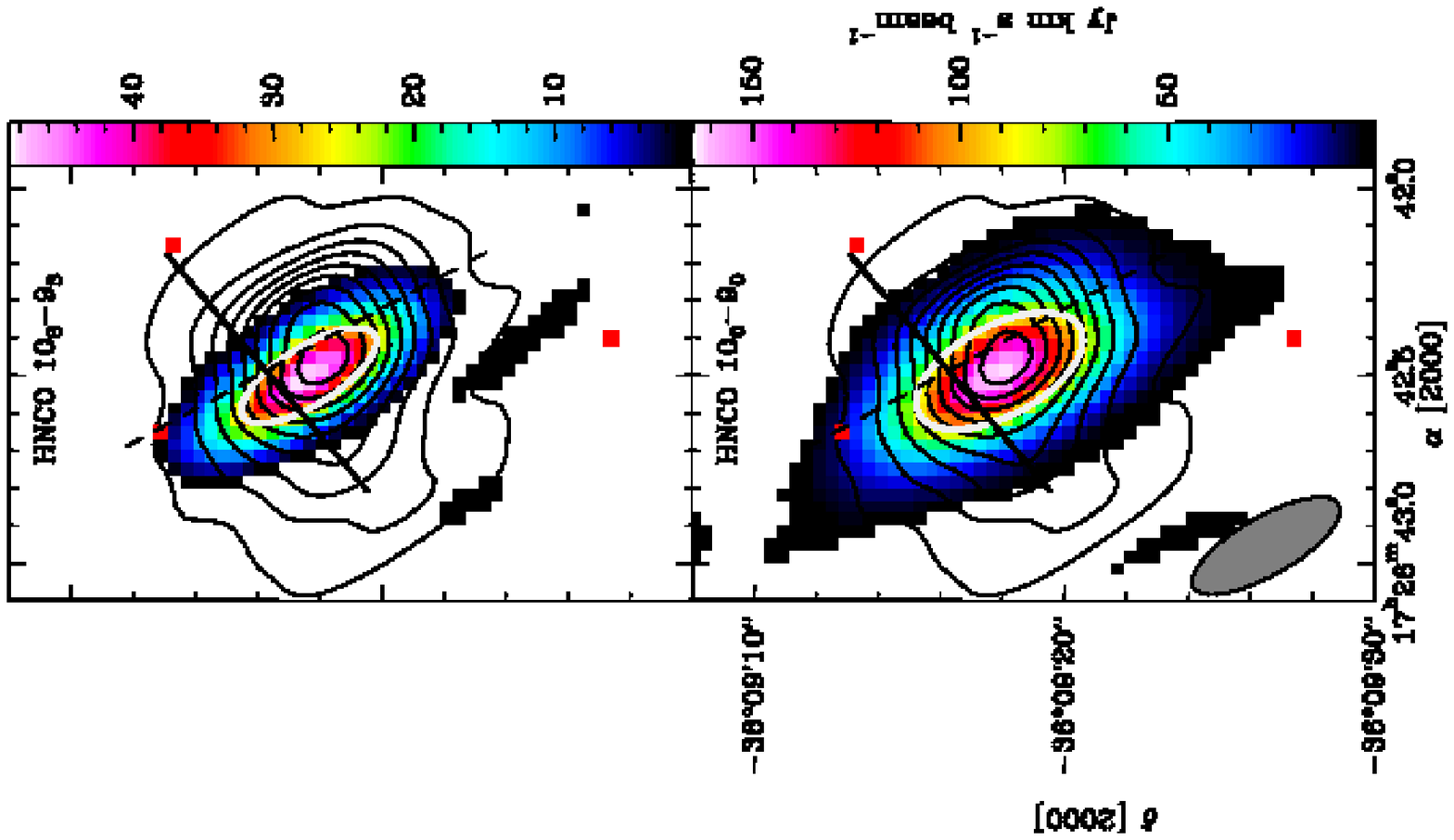}
\caption{Integrated emission of selected transitions of HNCO  in the region above the  
5$\sigma$ detection in the line data cubes ($\sim$ 0.5 Jy~beam$^{-1}$).
The black solid contours show the 1.3~mm continuum emission. 
Symbols are as in Fig.~\ref{ch3ohmom0}.\label{hncomom0}}
\end{figure}

\begin{figure}
\centering
\includegraphics[angle=-90,width=8cm]{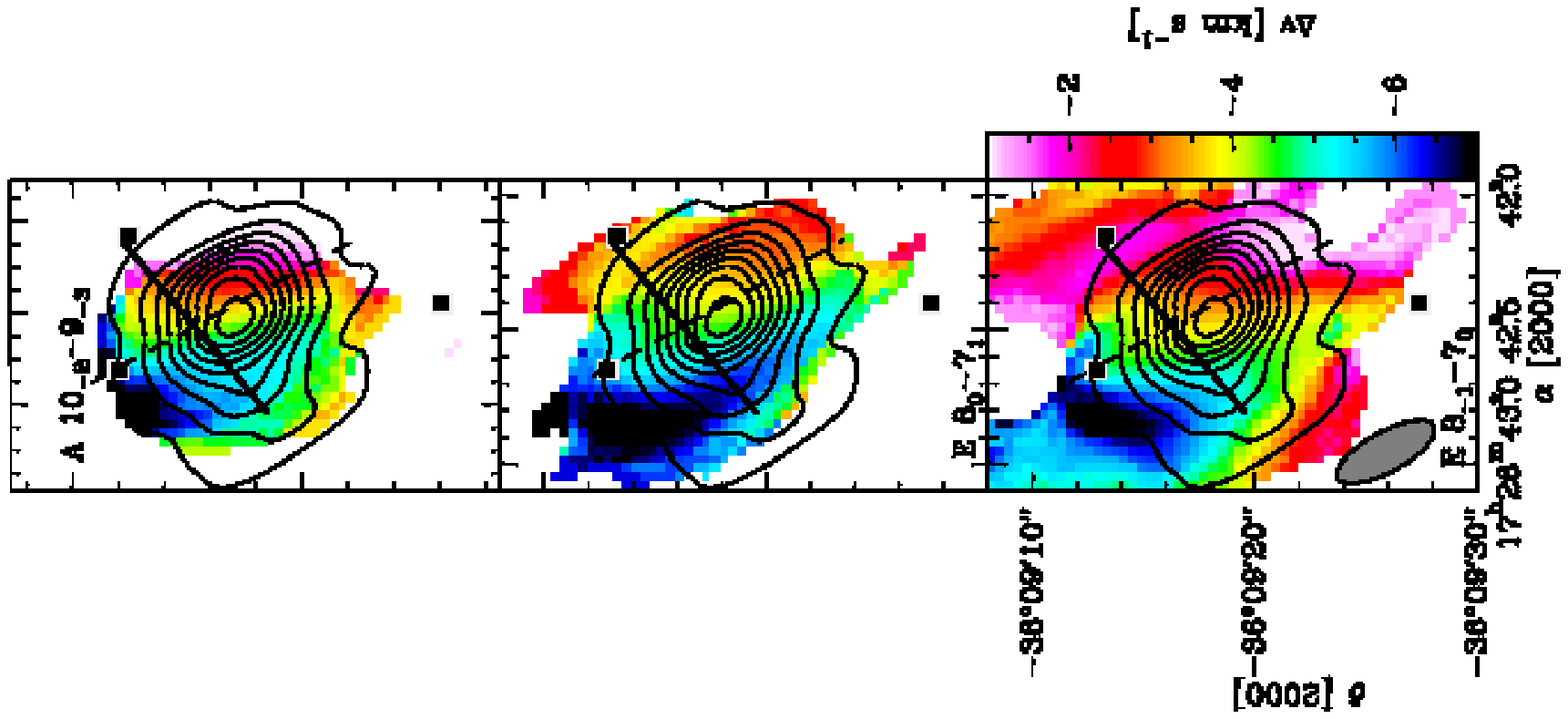}
\caption{First moment maps of selected transitions of CH$_3$OH. 
The black solid contours show the 1.3~mm continuum emission. 
Black squares mark the B1, R1 and R2 positions of Fig.~\ref{jets}. The other symbols are as in Fig.~\ref{ch3ohmom0}. The systemic velocity of the source lies between  
$-3.4$ and $-2.5$~km~s$^{-1}$ (see Sect.~\ref{mom1}). 
\label{ch3ohmom1}}
\end{figure}

As a first step to investigate the association of given transitions
with the molecular outflows in the region, with potential accretion
disks or with the envelope around protostars, we derived the
integrated intensity of different lines to study their spatial
distribution. In Figs.~\ref{ch3ohmom0},~\ref{ocsmom0}, \ref{ch3cnmom0}, and \ref{hncomom0} 
 we present the integrated intensity distribution of
 selected transitions (Table~\ref{id}). 
All the maps are elongated along the P.A. of
the beam of the observations. For all maps, we performed a 2-D Gaussian fit to infer
the  FWHM size and peak position of the integrated intensity distributions. The results are listed in Table~\ref{fit}.
All emission maps are resolved with our current resolution, 
except those of the HNCO $(10_3-9_3$), HC$_3$N $(24_{-2}-23_2)~v_7=2$,  
CH$_3^{13}$CN $(12_2-11_2$), and CH$_3$CN $(12_2-11_2)~v_8=1$ lines.
The fit suggests that the peak of the molecular emission does not coincide with the
1.3 mm continuum maximum, but it is located slightly to its NE, with a mean offset of $(9\farcs0,-17\farcs3)$ from the centre of the maps, $(0\farcs6,0\farcs3)$ from the peak of the mm continuum emission.
This could indicate that the hot core is associated with the  compact cm source VLA-2a identified
 by \citet{2008AJ....136.1455Z} (also shifted to the east of the mm continuum), whereas the bulk of continuum is
tracing colder material.  Opacity effects could play a role in this displacements. Actually, the CH$_3$CN
spectral pattern shows similar brightness temperatures for lines with $K \le$ 5 (Fig.~\ref{spectra-lsb}), suggesting
non negligible opacities. However, the distribution of less abundant isotopologues 
(see O$^{13}$CS) or vibrationally excited transitions
(HC$_3$N v$_{\rm 7}$ = 2 lines) 
seem to peak closer to the continuum peak, but still shifted to the east, near VLA-2a.

Methanol shows  extended emission along the axis of the OF2--OF3 outflows, and an elongated structure 
towards NW, i.e. in
the direction of the OF1 red-shifted outflow  
detected in H$_2$, extremely high velocity CO
and SO (see Fig.~\ref{jets}) and indicated by an arrow in
Figs.~\ref{ch3ohmom0}-\ref{hncomom0}. 
Interestingly, the occurrence of CH$_3$OH emission along the OF1 axis is more
evident  in the two lower excitation transitions (reported in Table~\ref{id}). 
As shown by the spectra in Fig.~\ref{spectra-lsb},
this is not due to  low dynamical range, since all methanol lines used in this analysis have a  signal-to-noise ratio 
larger than 40 at the peak of the continuum emission.

Also the
 OCS emission, and to some extent O$^{13}$CS emission, is elongated in the direction
the OF2--OF3 outflows, and shows a wisp of emission in the direction of OF1 (Fig.~\ref{ocsmom0}).

 Figure~\ref{ch3cnmom0} shows the zero-th moment
  maps derived from different lines of CH$_3$CN.  The
 spatial distribution of the $12_4-11_4$ line, the lowest  CH$_3$CN  excitation
 line used in our analysis,  extends in the
 direction of the OF1 jet and along the axis of the OF2--OF3
 outflows, while the other transitions are more compact. Also in this case we can exclude a bias due to different
 signal-to-noise ratios for the different CH$_3$CN lines.

The C$_2$H$_5$CN and HNCO ($10_0-9_0$) line maps are also elongated along the direction
of the OF2--OF3 axes, but their emission is more compact than for the other species discussed above.

In conclusion, the spatial
distribution of CH$_3$OH, OCS, and CH$_3$CN emission could be affected by the 
presence of the outflows. In particular, for CH$_3$OH and CH$_3$CN we find,
besides an elongation along the OF2--OF3 axes, also a good 
correlation with the direction of the OF1 outflow. 
Emission at the position R1, the peak of the CO ($2-1$) red-shifted emission at extremely high velocity 
in OF1, is detected from the CH$_3$OH ($8_{-1}-7_0$) and ($8_{0}-7_1$) lines and the CH$_3$CN ($12_4-11_4$) transition. 
In Sect.~\ref{mom1} we will analyse the first moment maps to investigate the  kinematics of the gas and better constrain 
the association of molecular emission with the molecular outflows from the IRAS\,17233 cluster.

\subsection{Velocity distribution}\label{mom1}

For  transitions analysed in Sect.~\ref{mom0}
  with emission   larger than 3 beam widths along the minor axis of the beam, 
we also computed the first moment distribution. This excludes  the HC$_3$N transitions from the analysis.

The detected velocity shifts are of the  
order of $\Delta \rm{v} \sim$ 6~km~s$^{-1}$, 
i.e. values larger than velocity 
resolution of the data cubes (0.5 km s$^{-1}$). 
All CH$_3$CN maps are blue-shifted compared to the   systemic velocity
($V_{\rm sys}$) of --3.4 km s$^{-1}$ \citep{1996A&AS..115...81B}. 
However, there is a large uncertainty in the determination of  $V_{\rm sys}$  
(between $-3.7$ and $-2.5$~km~s$^{-1}$ from single dish N$_2$H$^+$ and C$^{17}$O observations, \citealt{17233_irdc}), 
 probably due to the outflow multiplicity in the region.

As suggested by the analysis of the integrated intensity
distributions, CH$_3$OH clearly presents a velocity
gradient along the direction of the OF2--OF3 outflows, with
blue-shifted emission to the north-east and red-shifted emission to the
south-west.  Additionally, the  CH$_3$OH ($8_{-1}-7_0$) and ($8_0-7_1$)-$E$ lines 
show   red-shifted emission associated with the red lobe
of the OF1 jet (see Fig.~\ref{zoom}). A wisp of red-shifted emission towards OF1 is detected also in the 
 CH$_3$OH ($10_{-2}-9_{-3}$) line although marginally.

\begin{figure*}
\centering
\includegraphics[angle=-90,width=16cm]{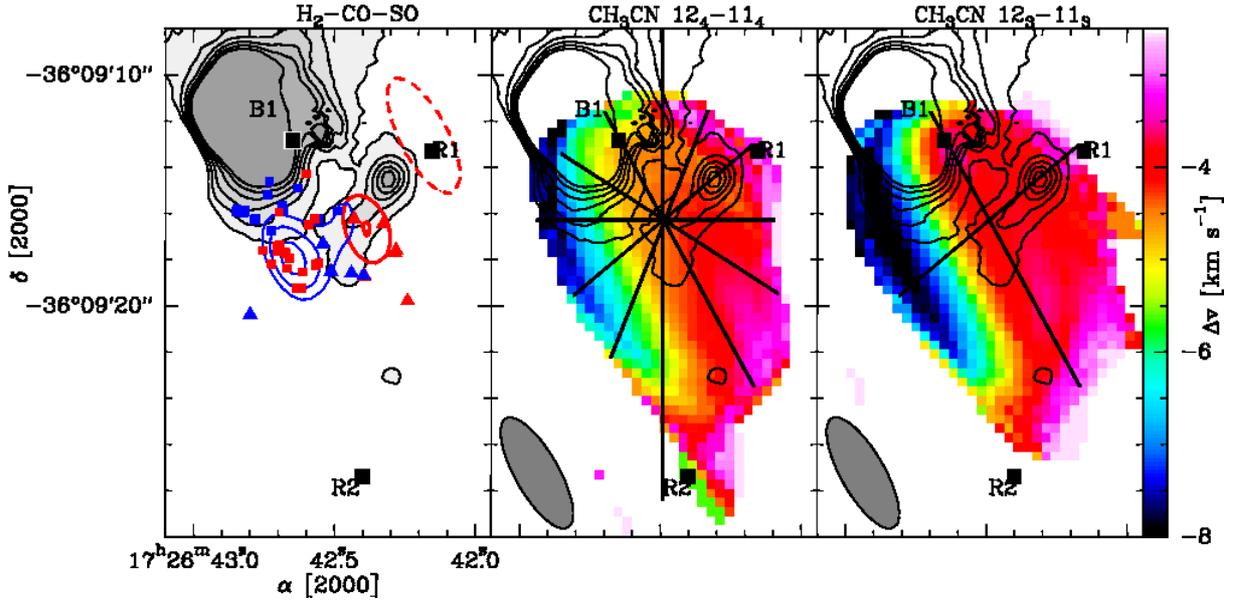}
\caption{Summary of the high angular resolution observations available for the
IRAS\,17233 cluster (see previous figures for details and references). 
{\it Left:} Maps of the H$_2$ emission (grey and solid contour levels). 
The  blue and red contours represent the SO$(5_6-4_5)$ blue- and red-shifted emission; the dashed red contour 
is the CO($2-1$) red-shifted emission at extremely high velocities. 
Overlaid on the map
are also the positions of the blue and red OH (squares) and H$_2$O (triangles) masers. 
{\it Middle:} 
Overlay of the CH$_3$CN(12$_4$--11$_4$) first moment map
(colour scale) with the H$_2$ emission map (black contours).
The black lines represent the directions used in the position-velocity plots of Fig.~\ref{pvs} and
are: the OF1 axis; the OF2--OF3 axis (NE--SW line); 
the N--S and E--W directions; the cuts with position angles of -20 and 60$^\circ$.
{\it Right:} Overlay of the CH$_3$CN(12$_3$--11$_3$) first moment map
(colour scale) with the H$_2$ emission map (black contours).  The black lines represent
the axes of the OF1 and OF2--OF3 outflows.  
The R1, R2 and B1 positions in OF1 and OF2--OF3 respectively are marked by  black squares.}
\label{summary}
\end{figure*}

\begin{figure}
\centering
\includegraphics[width=8cm,angle=-90]{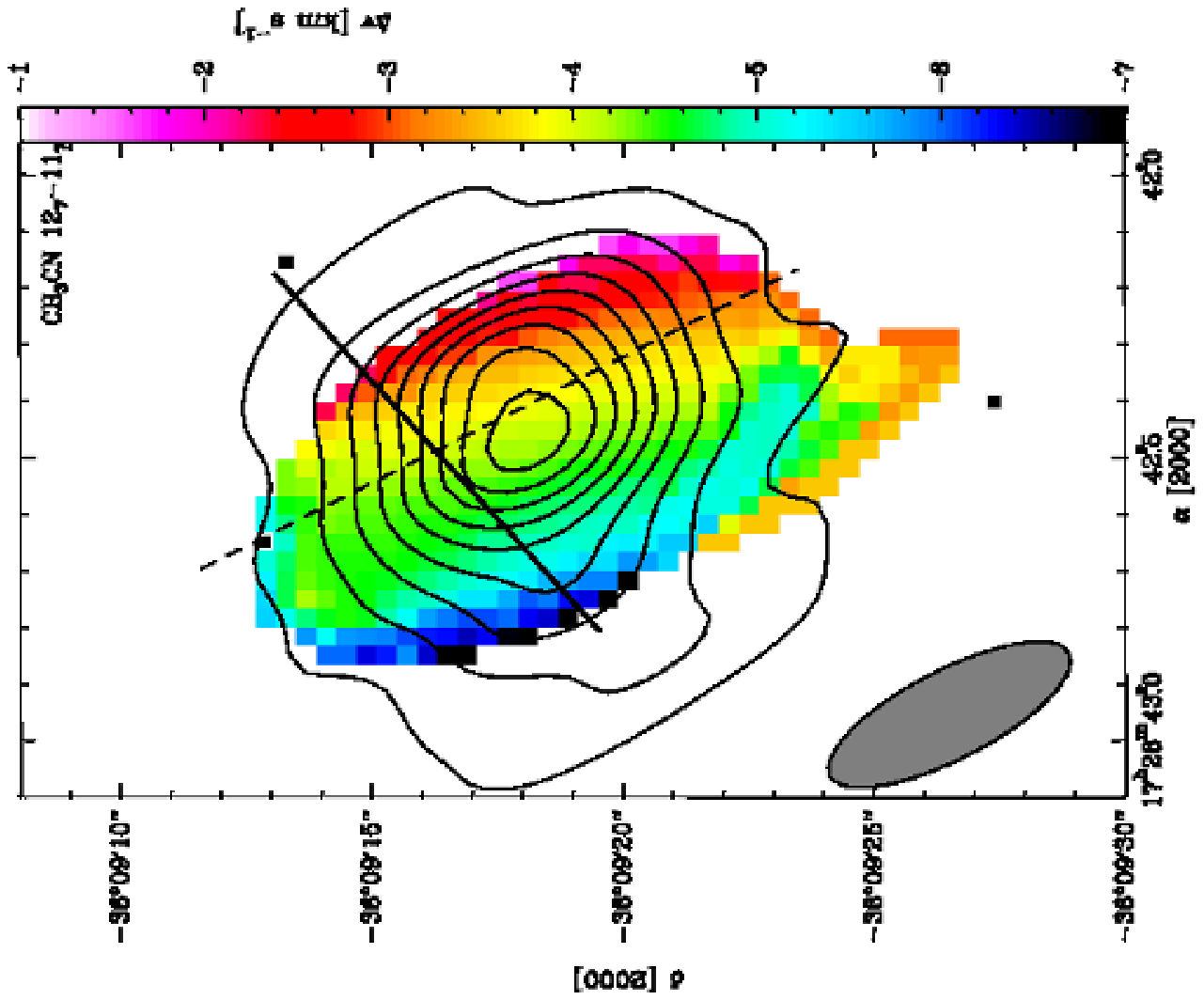}
\caption{First moment map of the 
 CH$_3$CN $(12_7-11_7)$ line. The black solid contours show the 1.3~mm continuum emission. 
Black squares mark the B1, R1 and R2 positions of Fig.~\ref{jets}. The other symbols are as in Fig.~\ref{ch3ohmom0}. The systemic velocity of the source lies between
$-3.4$ and $-2.5$~km~s$^{-1}$ (see Sect.~\ref{mom1}).
\label{ch3cnmom1}}
\end{figure}

For simplicity, we present the first moment distribution of
  the CH$_3$CN ($12_4-11_4$) line in Fig.~\ref{summary} while the
  CH$_3$CN ($12_7-11_7$)  transition is presented in Fig.~\ref{ch3cnmom1}. 
  We see the same behaviour seen in the CH$_3$OH ($10_{-2}-9_{-3}$) line in the
  ($12_4-11_4$) transition, the lowest energy analysed line of methyl
  cyanide: a linear velocity gradient along the E--W axis plus a
  red-shifted wisp of gas extending to the north-west towards the
  red-shifted lobe of the OF1 outflow.  The  CH$_3$CN ($12_7-11_7$) line has
  a more compact emission, and shows a linear velocity gradient along
  the E--W direction, almost perpendicular to the OF2--OF3 axis. Although not shown here, 
  the velocity field of the CH$_3$CN ($12_8-11_8$) line agrees with that of the CH$_3$CN ($12_7-11_7$) transition.

In
  order to further investigate the association of the CH$_3$CN
  ($12_4-11_4$) line with the OF1 outflow, in Fig.~\ref{summary} we
  compare its first moment distribution to the H$_2$
  distribution and to other tracers of kinematics in the region. 
  The SO blue- and red-shifted
  emission and the blue- and red-shifted H$_2$O masers trace the blue and red- shifted
lobes of the OF1
  outflow (see Fig.~\ref{summary} and \citealt{2009A&A...507.1443L}).  
If we take into account (i) the
  EHV OF1 red-lobe in CO ($2-1$), (ii) the SO red- and blue-shifted
  emission, and (iii) the water maser spots, then the contribution of
  OF1 to the kinematics of the CH$_3$CN ($12_4-11_4$) transition appears
  reasonably evident. Furthermore, Fig.~\ref{summary} also shows
  red-shifted emission elongated towards the R2 red lobe of OF2--OF3
  and blue-shifted emission close to the B1 peak of OF2--OF3 as
  observed in CO (see Fig.~\ref{overview}). In
  Fig.~\ref{summary} we also show the velocity field of the CH$_3$CN
  ($12_3-11_3$) line which has been excluded from our analysis because 
of weak blending with other features towards the centre position.
However, being lower in energy than the   ($12_4-11_4$)  transition,
the ($12_3-11_3$) line has a larger emitting size and can help the interpretation of
the kinematics of the other CH$_3$CN transitions. 
In this case, the N--W
 velocity gradient seen 
in the high J CH$_3$CN transitions and to some extent also in the ($12_4-11_4$) line is less pronounced,
while red-shifted emission is clearly detected towards the R1 position in OF1 and towards the red-shifted
lobe of the OF2--OF3 outflows.

\begin{figure}
\centering
\includegraphics[angle=-90,width=7cm]{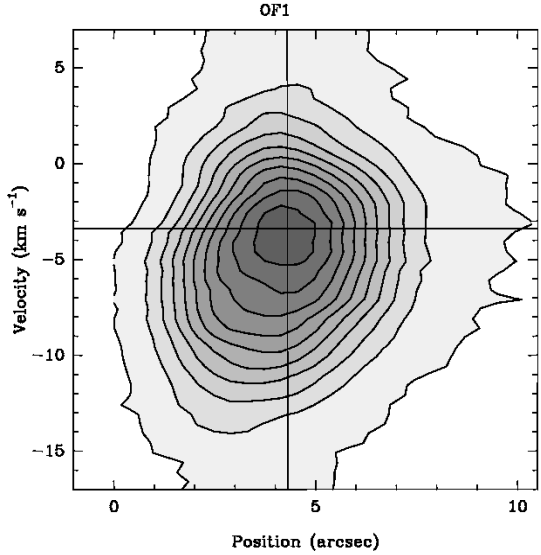}
\caption{
Position-velocity plot of the CH$_3$CN(12$_4-11_4$) line along the
OF1 axis. Contour levels range from 5$\sigma$ (0.5~Jy~beam$^{-1}$) in
step of 20$\sigma$. The horizontal line marks the systemic velocity of $-3.4$~km~s$^{-1}$, the vertical line 
the crossing point of all slits.}\label{pv}
\end{figure}

\begin{figure}
\centering
\includegraphics[angle=-90,width=8cm]{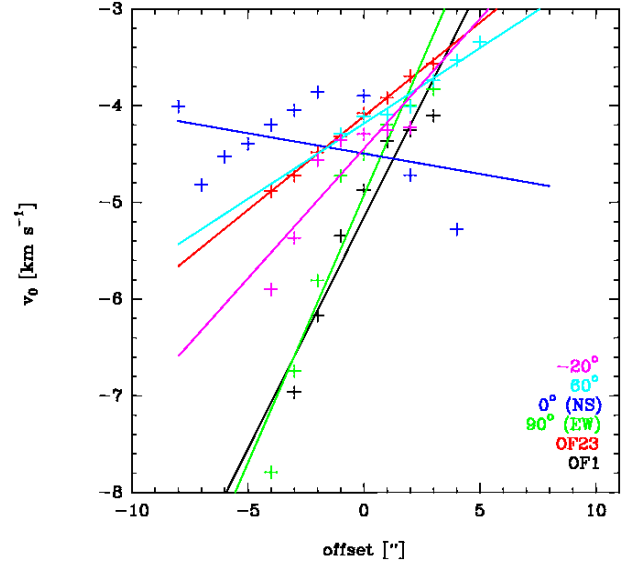}
\caption{Peak velocity of the 
 of CH$_3$CN(12$_4$--11$_4$) line along six different directions outlined in Fig.~\ref{summary} 
as function of the offset position along the slit.
Different colours denote different slits. The solid lines represent the fit to the data.}
\label{pvs}
\end{figure}

As an additional check,
  we derived position-velocity plots  along several slits placed to sample all 
possible directions: the OF1 axis (P.A. $-50^\circ$), the NE--SW direction (P.A. 30$^\circ$, as representative
of the three possible directions of the OF2--OF3 outflows), two slits with position angles 60$^\circ$ and -20$^\circ$, 
the N--S and E--W directions. All six slits are shown in
  Fig.~\ref{summary}. In Fig.~\ref{pv} we show the PV diagram along the OF1 axis which shows 
that CH$_3$CN ($12_4-11_4$) emission 
is spread over a broad  range of velocities ($\sim 15$~km s$^{-1}$).
We then extracted spectra of
  the CH$_3$CN $(12_4-11_4)$ transition for each slit, and fitted the line with a Gaussian profile. For each
  slit, we used a linear fit to reproduce the peak velocity as
  function of the offset position along the slit. 
 Results are shown  in Fig.~\ref{pvs}. 
The largest velocity gradients are detected along 
the E--W direction ($\sim114$~km~s$^{-1}$~pc$^{-1}$) and the OF1 jet ($\sim99$~km~s$^{-1}$~pc$^{-1}$). 
The gradients along OF2--OF3, and the slits at 60$^\circ$ and $-20^\circ$ have  
similar
gradients ($32-55$~km~s$^{-1}$~pc$^{-1}$). 
Only the velocity gradient along the N--S direction is negligible, and no  fit could be performed for this slit.

 Clearly, the velocity gradient along OF1 plays an important role in the kinematics of the inner gas around the protostars.
 This
strongly supports our interpretation that the velocity field of the CH$_3$CN $(12_4-11_4)$ line 
is affected by the OF1 outflow.
 In Sect.~\ref{discussion}, we will discuss different scenarios for the velocity gradient
detected along the E--W axis.

\begin{figure}
\centering
\includegraphics[width=8cm,angle=-90]{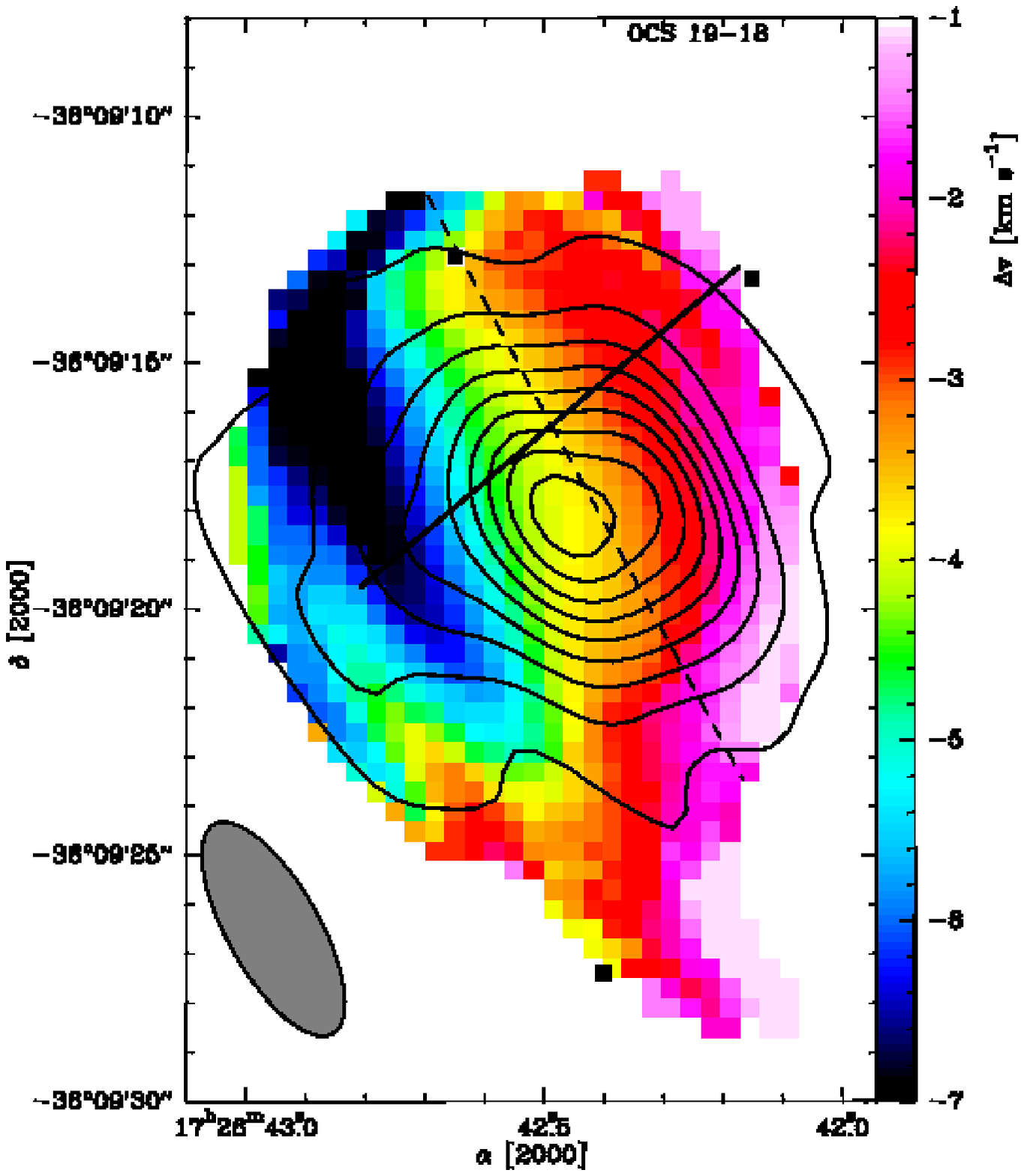}
\caption{First moment map of the OCS ($19-18$) transition. 
The black solid contours show the 1.3~mm continuum emission. 
Black squares mark the B1, R1 and R2 positions of Fig.~\ref{jets}. The other symbols are as in Fig.~\ref{ch3ohmom0}. The systemic velocity of the source lies between
$-3.4$ and $-2.5$~km~s$^{-1}$ (see Sect.~\ref{mom1}).
\label{ocsmom1}}
\end{figure}

\begin{figure}
\centering
\includegraphics[width=8cm,angle=-90]{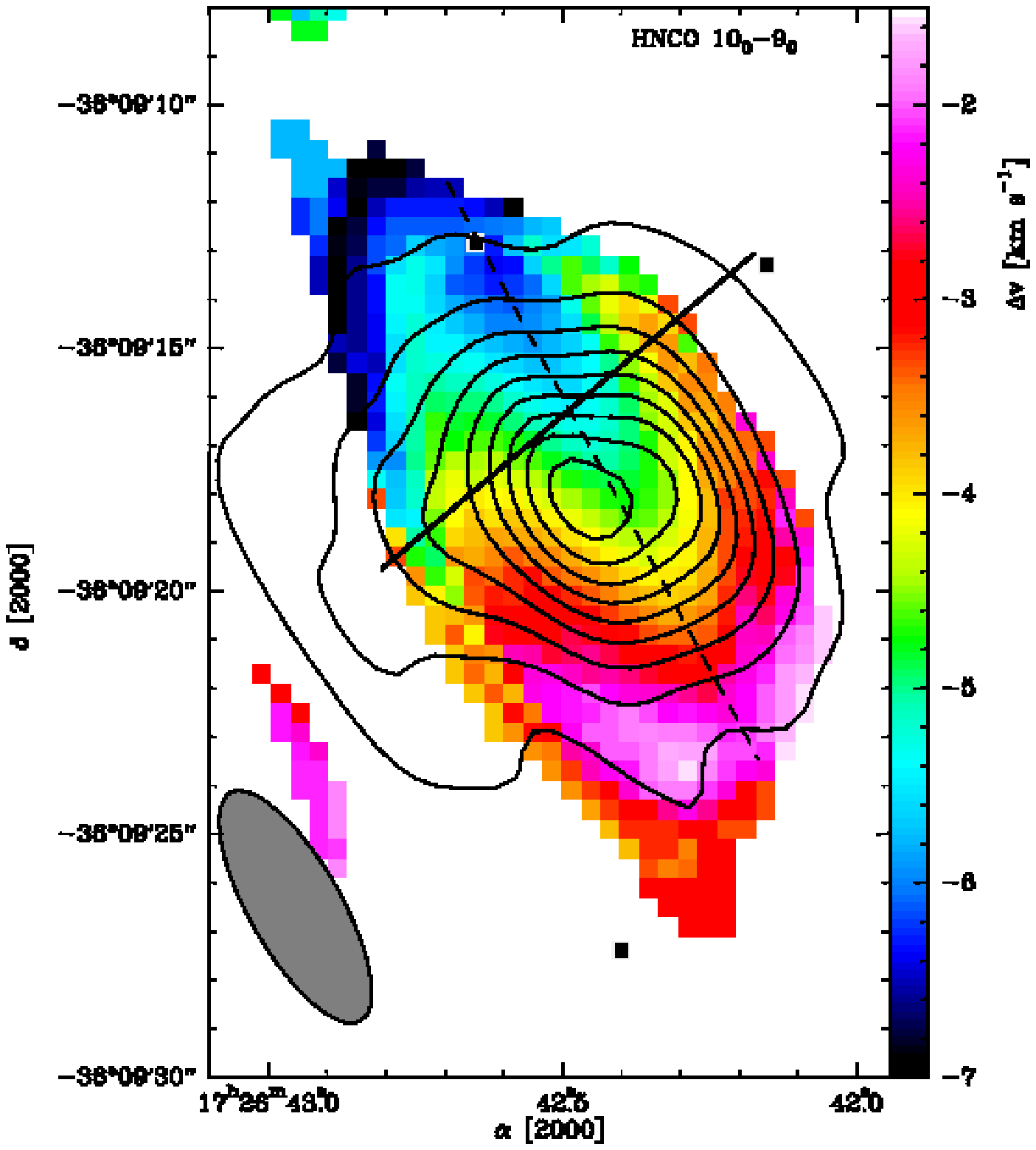}
\caption{First moment maps of the HNCO ($10_0-9_0$) line.  The black solid contours show the 1.3~mm continuum emission. 
Black squares mark the B1, R1 and R2 positions of Fig.~\ref{jets}.  The other symbols are as in Fig.~\ref{ch3ohmom0}. The systemic velocity of the source lies between
$-3.4$ and $-2.5$~km~s$^{-1}$ (see Sect.~\ref{mom1}).
\label{hncomom1}}
\end{figure}

The velocity field of the  OCS ($19-18$) transition  (Fig.~\ref{ocsmom1})
is  consistent with those 
of CH$_3$OH and of the CH$_3$CN $(12_4-11_4)$ transition and shows red-shifted emission towards the red peak R1 of OF1 and 
R2 of OF2--OF3 and blue-shifted emission close to one of the blue peaks of OF2--OF3, B1. 
O$^{13}$CS (not shown here) is  consistent with the picture depicted by high J CH$_3$CN transitions.

Finally, the HNCO emission shows a velocity gradient in the NE--SW direction along the axis of OF2--OF3
(see Fig.~\ref{hncomom1}). The first moment distribution of the C$_2$H$_5$CN (not shown here) is consistent 
with those of the high J CH$_3$CN lines.

\section{Discussion}\label{discussion}

As seen in Sect.~\ref{mom1}, different transitions trace different kinematics. On the base of their first moment maps
we can distinguish three cases:

\begin{enumerate}

\item transitions with 
 a velocity gradient along the NE--SW direction (OF2--OF3), and 
additionally red-shifted emission towards NW (OF1). This is seen in CH$_3$OH, CH$_3$CN ($12_4-11_4$), and OCS;

\item transitions with a velocity gradient along the NE--SW axis (OF2--OF3) (HNCO);

\item  transitions with a velocity gradient along the  E--W axis (high excitation CH$_3$CN lines).

\end{enumerate}

As seen before, the first case can be explained by a
combination of the blue- and red-shifted lobes of the OF1, OF2--OF3 outflows,
indicating for CH$_3$OH, CH$_3$CN($12_4-11_4$), and OCS 
 at least a non negligible contamination from 
the multiple outflows in the IRAS 17233--3606 cluster.

While the association of CH$_3$OH and OCS with molecular outflows  
is not surprising since they have been detected in several chemical
active molecular outflows \citep[e.g.,][]{1997ApJ...487L..93B}, 
the detection of a similar behaviour in the CH$_3$CN ($12_4-11_4$) transition
is unexpected. CH$_3$CN is usually
considered a pure hot core tracer and is often used to
investigate velocity fields around massive YSOs to search for evidence of rotation 
(see \citealt{1999A&A...345..949C,2004ApJ...601L.187B,2005A&A...435..901B,2011A&A...525A.151B}, but  also discussion in 
\citealt{2004ApJ...616..301G} and \citealt{2008ApJ...675..420A}). However, in the last years
\citet{2008ApJ...681L..21A} and \citet{2009A&A...507L..25C} detected
CH$_3$CN towards the molecular outflow driven by the low-mass YSO L1157-mm, hence showing that
also this species  can be altered by the
passage of bow shocks, although with
abundances much lower than in hot cores and hot corinos around high- and low-mass YSOs
\citep[e.g., ][]{1993A&A...276..489O,2005A&A...435..901B,2007A&A...463..601B}. 
Additionally, \citet{2010arXiv1009.1426Z} recently proposed that CH$_3$CN  emission in Orion-KL is also excited by shocks. 
In the present case, 
 the velocity gradient detected in CH$_3$CN could be due to outflow-envelope interactions as already seen in low-mass star class 0 sources \citep{2004A&A...416..631B,2005ApJ...624..232A,2005ApJ...619..948L} in other molecular species.

The second kinematical pattern (HNCO) can be easily associated with the motions
of the OF2--OF3 outflows. HNCO is abundant in densest regions of
molecular clouds  \citep{1984ApJ...280..608J,1999ApJ...514L..43W} as well as in  regions of low
velocity shocks \citep{2008ApJ...678..245M}.
However, very recently also HNCO has been detected towards the chemical active L1157 flow by
\citet{2010A&A...516A..98R}, showing that the HNCO line profiles exhibit the same
characteristics as other well-known tracers
of shocked gas and supporting the idea that HNCO is a good tracer
of interstellar shocks.
Our data support this scenario,
reporting the first detection of HNCO emission in outflows driven by massive YSOs.

The third case (velocity gradient almost perpendicular to OF2--OF3) 
is the most intriguing since CH$_3$CN emission has been extensively used in the last years to
trace rotating toroids
supposed to host accretion disks around massive YSOs. 
This has been used to support  a mechanism for the formation of massive stars similar to that of  low-mass stars.
 In the case of IRAS\,17233, at least three possible interpretations can be given for 
the E--W velocity gradient detected in
the CH$_3$CN lines: rotation, contamination by outflows, velocity differences between cores.

In the first case, the E--W velocity gradient
could trace a rotating structure perpendicular to OF2--OF3, 
the dominant outflow of this region. In this case, following \citet{2004ApJ...601L.187B,2005A&A...435..901B} 
we computed the dynamical mass $M_{dyn}$  
assuming equilibrium between centrifugal and gravitational forces. We took as
rotation velocity $V_{rot}=3$ km~s$^{-1}$, value corresponding to half the velocity range measured in 
the CH$_3$CN $(12_4-11_4)$ line and as radius half the deconvolved FWHM size of the continuum emission ($S=5\farcs3$, 
see Table~\ref{fit}).
This corresponds  to a dynamical mass of 27~M$_\odot$ for an edge-on toroid. This is likely a lower limit 
to the real dynamical mass since the  inclination angle is unknown. 
Such value is  consistent
with those estimated by \citet{2004ApJ...601L.187B,2011A&A...525A.151B} for other sources.

However, contamination by outflows 
questions this interpretation, since 
non negligible velocity gradients are detected 
along the OF2--OF3 axis and along other directions (see Sect.~\ref{mom1} and Fig.~\ref{pvs}). In this context, the velocity gradient along the E-W
direction could be due to a composition of motions along OF2--OF3 and OF1. 
Finally, since four cm compact VLA-sources are in the beam of our SMA data, and they are roughly located along
the E-W axis,  
 one could consider that the kinematical patterns here reported 
are simply due to a velocity difference of the sources.

Clearly, higher angular observations
are needed to discriminate between these interpretations. Given its low declination, ALMA 
is the ideal instrument to investigate the source at higher resolution.

\section{Implications for studies of high-mass YSOs}\label{implications}

Although not conclusive for the search of tracers of rotation around
massive YSOs, we believe that our study of IRAS\,17233 has important
implications for searches of disks in high-mass star forming regions.
IRAS\,17233 is an exceptionally close and active site of  massive
star formation. Therefore, even observations with a limited angular
resolution allow us to resolve small linear scales. 
Since the distance of the source is $D\le 1$~kpc, our beam of $\sim 5.4\arcsec\times 2\arcsec$ corresponds to 
a maximum linear scale of 5400 $\times$ 2000~AU, thus comparable with state-of-the-art 
interferometric  observations of more distant massive YSOs. 
For example, 
the recent study of \citet{2011A&A...525A.151B} was performed 
with an linear resolution between 2400 and
50\,000 AU.

\begin{figure}
\centering
\includegraphics[width=9cm,angle=-90]{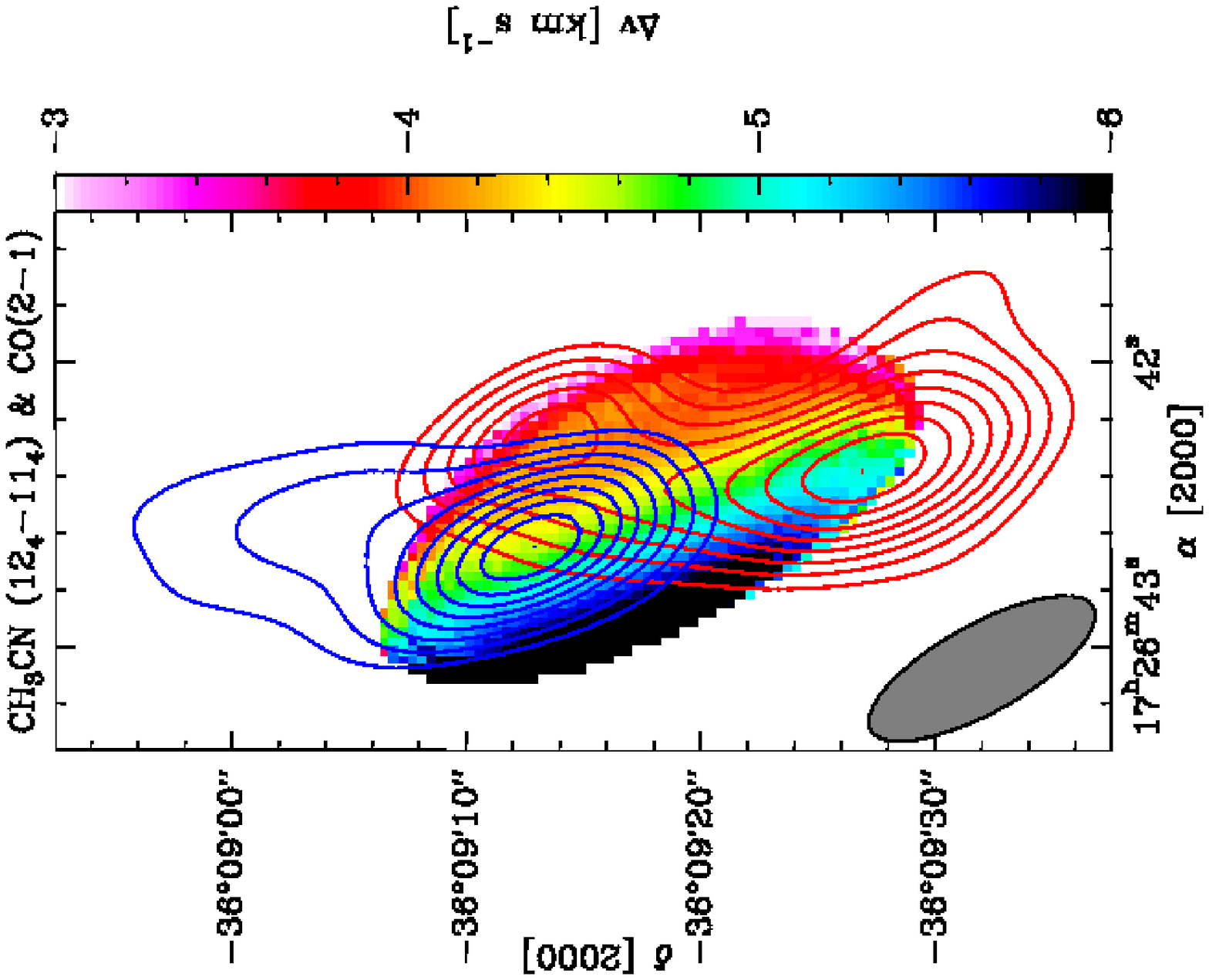}
\caption{Velocity gradient of the CH$_3$CN (12$_4-11_4$) 
transition smoothed to an angular resolution of $10\farcs86\times 3\farcs78$. 
The red and blue solid contours show the red- and blue-shifted CO(2--1) emission 
smoothed to the same angular resolution.}\label{smooth}
\end{figure}

To verify whether the detection of OF1 in the CH$_3$CN ($12_4-11_4$)
transition would have been possible with lower linear resolution
(i.e., a more distant object), we smoothed our data cube of a factor
of two in spatial resolution.  This corresponds to observations of a
massive YSO at a distance twice as large as that of IRAS\,17233, but
with the same angular resolution as our observations of
IRAS\,17233. We then computed moment maps for the CH$_3$CN
($12_4-11_4$) transition following the same procedure described in
Sect.~\ref{res}. We also smoothed the CO ($2-1$) data cube to the same
angular resolution, and derived integrated intensity maps for the
blue- and red-shifted emission in the same velocity range as in
Fig.~\ref{overview}.  The resulting maps are presented in
Fig.~\ref{smooth}.

The data at lower angular
resolution show a compact region detected in CH$_3$CN with a velocity
gradient perpendicular to the dominant outflow of the region, OF2--OF3.
 Clearly, the red-shifted emission along the OF1
axis is not longer detected, and the derived velocity field is very
similar to that detected for the CH$_3$CN ($K=7,8$) lines. In other words 
the picture of Fig.~\ref{smooth} could easily be explained with a pure velocity
gradient perpendicular to the N--S outflow as seen in CO.
We note that for sources at distances $\ge$ 10~kpc also the detection of H$_2$ is challenging.
Indeed, in the case of IRAS\,17233, the H$_2$ data  were fundamental to reveal OF1, since 
H$_2$  gives a much better picture of the outflow multiplicity  than CO because
 it traces hot jets and is not  
contaminated by infalling
envelopes and/or swept up cavities.

\section{Conclusions}

In this study, we analysed  interferometric line observations of the massive star forming region 
IRAS\,17233 at 230~GHz with an angular resolution of $5\farcs4\times 1\farcs9$. 
We studied the spatial
distribution and velocity field of different lines and compared them to
the distribution of outflows in the region to investigate the nature of the emission.
We report the first detection of HNCO in molecular outflows from  massive YSOs. 
 Although other interpretations cannot
 be ruled out, we find that the velocity gradient observed in CH$_3$CN could be
 significantly affected by the OF1 outflow. This follows
the recent detection of both molecules in the chemically active outflow L1157, powered
by a low-mass YSO. 
The possible detection of CH$_3$CN in molecular outflows 
questions the choice of the tracer to probe rotating structures around YSOs: one has to keep in
 mind that the same molecule could be tracing outflows in some objects and
 rotating disks/toroids in others.
Further studies of high-mass star forming regions are needed to verify whether IRAS 17233--3606 represents
a standard laboratory where to investigate star formation or whether it is a special case
 with a too complex scenario.
The present observations, with the possible association of CH$_3$CN with
the OF1 outflow in a few thousands AU around the peak
of the continuum, 
stress   the importance of  (i) very high-spatial resolution, less than 1000 AU, 
and of (ii)  H$_2$ images 
for detailed studies of kinematics in regions as those around massive YSOs.
Assuming typical distances of massive star forming regions between 5 and 10~kpc, 
a resolution of 1000~AU corresponds to an angular size of $0\farcs1$--$0\farcs2$. 
Even smaller angular resolutions are then needed
to study in details the kinematics of such regions. 
Only studies of massive YSOs with angular resolution 
$\le 0\farcs1$ (for distances $>5$~kpc) will clarify 
whether or not IRAS\,17233 represents a special case and will shed 
light on to the choice of the molecular tracers for rotating structures.

As a final remark, given the chemical differences in hot cores, and
contamination by outflow emission, potentially in all species, it well
may be the case that no molecular species or transition will work for
every source under every circumstance.  However, we can try to
formulate criteria for convincing detection of a rotating disk/toroid in millimeter molecular emission: (i)
such a detection should be based on the analysis of high excitation multiple lines and
multiple species; (ii) the lines used in the analysis should be
unblended; (iii) any possible contribution from outflows should be
ruled out; (iv) all lines should be consistent with a single velocity
field of a rotating body.  As noted in the previous section, selective
tracers of  hot jets as H$_2$, or SiO, are better suited to understand
the multiplicity and morphology of outflows than CO and may therefore
better help to understand the kinematics of the cores.

\begin{acknowledgements}
We are grateful to M. Walmsley and A. Gusdorf for fruitful discussions and suggestions.  We would also like to 
thank an anonymous referee
for a careful review of the manuscript.    
\end{acknowledgements}


\begin{thebibliography}{}
\expandafter\ifx\csname natexlab\endcsname\relax\def\natexlab#1{#1}\fi

\bibitem[{{Araya} {et~al.}(2008){Araya}, {Hofner}, {Kurtz}, {Olmi}, \&
  {Linz}}]{2008ApJ...675..420A}
{Araya}, E., {Hofner}, P., {Kurtz}, S., {Olmi}, L., \& {Linz}, H. 2008, \apj,
  675, 420

\bibitem[{{Arce} {et~al.}(2008){Arce}, {Santiago-Garc{\'{\i}}a},
  {J{\o}rgensen}, {Tafalla}, \& {Bachiller}}]{2008ApJ...681L..21A}
{Arce}, H.~G., {Santiago-Garc{\'{\i}}a}, J., {J{\o}rgensen}, J.~K., {Tafalla},
  M., \& {Bachiller}, R. 2008, \apjl, 681, L21

\bibitem[{{Arce} \& {Sargent}(2005)}]{2005ApJ...624..232A}
{Arce}, H.~G. \& {Sargent}, A.~I. 2005, \apj, 624, 232

\bibitem[{{Bachiller} \& {P{\'e}rez Guti{\'e}rrez}(1997)}]{1997ApJ...487L..93B}
{Bachiller}, R. \& {P{\'e}rez Guti{\'e}rrez}, M. 1997, \apjl, 487, L93

\bibitem[{{Beltr{\'a}n} {et~al.}(2011){Beltr{\'a}n}, {Cesaroni}, {Neri}, \&
  {Codella}}]{2011A&A...525A.151B}
{Beltr{\'a}n}, M.~T., {Cesaroni}, R., {Neri}, R., \& {Codella}, C. 2011, \aap,
  525, A151

\bibitem[{{Beltr{\'a}n} {et~al.}(2004{\natexlab{a}}){Beltr{\'a}n}, {Cesaroni},
  {Neri}, {Codella}, {Furuya}, {Testi}, \& {Olmi}}]{2004ApJ...601L.187B}
{Beltr{\'a}n}, M.~T., {Cesaroni}, R., {Neri}, R., {et~al.} 2004{\natexlab{a}},
  \apjl, 601, L187

\bibitem[{{Beltr{\'a}n} {et~al.}(2005){Beltr{\'a}n}, {Cesaroni}, {Neri},
  {Codella}, {Furuya}, {Testi}, \& {Olmi}}]{2005A&A...435..901B}
{Beltr{\'a}n}, M.~T., {Cesaroni}, R., {Neri}, R., {et~al.} 2005, \aap, 435, 901

\bibitem[{{Beltr{\'a}n} {et~al.}(2004{\natexlab{b}}){Beltr{\'a}n}, {Gueth},
  {Guilloteau}, \& {Dutrey}}]{2004A&A...416..631B}
{Beltr{\'a}n}, M.~T., {Gueth}, F., {Guilloteau}, S., \& {Dutrey}, A.
  2004{\natexlab{b}}, \aap, 416, 631

\bibitem[{{Beuther} {et~al.}(2004){Beuther}, {Hunter}, {Zhang}, {Sridharan},
  {Zhao}, {Sollins}, {Ho}, {Ohashi}, {Su}, {Lim}, \& {Liu}}]{18089_disk}
{Beuther}, H., {Hunter}, T.~R., {Zhang}, Q., {et~al.} 2004, \apjl, 616, L23

\bibitem[{{Beuther} {et~al.}(2002){Beuther}, {Schilke}, {Menten}, {Motte},
  {Sridharan}, \& {Wyrowski}}]{2002ApJ...566..945B}
{Beuther}, H., {Schilke}, P., {Menten}, K.~M., {et~al.} 2002, \apj, 566, 945

\bibitem[{{Beuther} {et~al.}(2005{\natexlab{a}}){Beuther}, {Schilke}, {Menten},
  {Motte}, {Sridharan}, \& {Wyrowski}}]{2005ApJ...633..535B}
{Beuther}, H., {Schilke}, P., {Menten}, K.~M., {et~al.} 2005{\natexlab{a}},
  \apj, 633, 535

\bibitem[{{Beuther} {et~al.}(2009){Beuther}, {Walsh}, \&
  {Longmore}}]{2009ApJS..184..366B}
{Beuther}, H., {Walsh}, A.~J., \& {Longmore}, S.~N. 2009, \apjs, 184, 366

\bibitem[{{Beuther} {et~al.}(2005{\natexlab{b}}){Beuther}, {Zhang},
  {Sridharan}, \& {Chen}}]{2005ApJ...628..800B}
{Beuther}, H., {Zhang}, Q., {Sridharan}, T.~K., \& {Chen}, Y.
  2005{\natexlab{b}}, \apj, 628, 800

\bibitem[{{Bonnell} {et~al.}(2007){Bonnell}, {Larson}, \&
  {Zinnecker}}]{2007prpl.conf..149B}
{Bonnell}, I.~A., {Larson}, R.~B., \& {Zinnecker}, H. 2007, Protostars and
  Planets V, 149

\bibitem[{{Bottinelli} {et~al.}(2007){Bottinelli}, {Ceccarelli}, {Williams}, \&
  {Lefloch}}]{2007A&A...463..601B}
{Bottinelli}, S., {Ceccarelli}, C., {Williams}, J.~P., \& {Lefloch}, B. 2007,
  \aap, 463, 601

\bibitem[{{Brand} \& {Blitz}(1993)}]{1993A&A...275...67B}
{Brand}, J. \& {Blitz}, L. 1993, \aap, 275, 67

\bibitem[{{Bronfman} {et~al.}(1996){Bronfman}, {Nyman}, \&
  {May}}]{1996A&AS..115...81B}
{Bronfman}, L., {Nyman}, L.-A., \& {May}, J. 1996, \aaps, 115, 81

\bibitem[{{Caswell} {et~al.}(1980){Caswell}, {Haynes}, \&
  {Phys}}]{1980IAUC.3509....2C}
{Caswell}, J.~L., {Haynes}, R.~F., \& {Phys}, J. 1980, \iaucirc, 3509, 2

\bibitem[{{Cesaroni} {et~al.}(1999){Cesaroni}, {Felli}, {Jenness}, {Neri},
  {Olmi}, {Robberto}, {Testi}, \& {Walmsley}}]{1999A&A...345..949C}
{Cesaroni}, R., {Felli}, M., {Jenness}, T., {et~al.} 1999, \aap, 345, 949

\bibitem[{{Cesaroni} {et~al.}(1997){Cesaroni}, {Felli}, {Testi}, {Walmsley}, \&
  {Olmi}}]{1997A&A...325..725C}
{Cesaroni}, R., {Felli}, M., {Testi}, L., {Walmsley}, C.~M., \& {Olmi}, L.
  1997, \aap, 325, 725

\bibitem[{{Cesaroni} {et~al.}(2007){Cesaroni}, {Galli}, {Lodato}, {Walmsley},
  \& {Zhang}}]{2007prpl.conf..197C}
{Cesaroni}, R., {Galli}, D., {Lodato}, G., {Walmsley}, C.~M., \& {Zhang}, Q.
  2007, Protostars and Planets V, 197

\bibitem[{{Codella} {et~al.}(2009){Codella}, {Benedettini}, {Beltr{\'a}n},
  {Gueth}, {Viti}, {Bachiller}, {Tafalla}, {Cabrit}, {Fuente}, \&
  {Lefloch}}]{2009A&A...507L..25C}
{Codella}, C., {Benedettini}, M., {Beltr{\'a}n}, M.~T., {et~al.} 2009, \aap,
  507, L25

\bibitem[{{Comito} {et~al.}(2005){Comito}, {Schilke}, {Phillips}, {Lis},
  {Motte}, \& {Mehringer}}]{2005ApJS..156..127C}
{Comito}, C., {Schilke}, P., {Phillips}, T.~G., {et~al.} 2005, \apjs, 156, 127

\bibitem[{{Fa{\'u}ndez} {et~al.}(2004){Fa{\'u}ndez}, {Bronfman}, {Garay},
  {Chini}, {Nyman}, \& {May}}]{2004A&A...426...97F}
{Fa{\'u}ndez}, S., {Bronfman}, L., {Garay}, G., {et~al.} 2004, \aap, 426, 97

\bibitem[{{Fish} {et~al.}(2005){Fish}, {Reid}, {Argon}, \&
  {Zheng}}]{2005ApJS..160..220F}
{Fish}, V.~L., {Reid}, M.~J., {Argon}, A.~L., \& {Zheng}, X.-W. 2005, \apjs,
  160, 220

\bibitem[{{Fix} {et~al.}(1982){Fix}, {Mutel}, {Gaume}, \&
  {Claussen}}]{1982ApJ...259..657F}
{Fix}, J.~D., {Mutel}, R.~L., {Gaume}, R.~A., \& {Claussen}, M.~J. 1982, \apj,
  259, 657

\bibitem[{{Forster} \& {Caswell}(1989)}]{1989A&A...213..339F}
{Forster}, J.~R. \& {Caswell}, J.~L. 1989, \aap, 213, 339

\bibitem[{{Gibb} {et~al.}(2004){Gibb}, {Wyrowski}, \&
  {Mundy}}]{2004ApJ...616..301G}
{Gibb}, A.~G., {Wyrowski}, F., \& {Mundy}, L.~G. 2004, \apj, 616, 301

\bibitem[{{Hughes} \& {MacLeod}(1993)}]{1993AJ....105.1495H}
{Hughes}, V.~A. \& {MacLeod}, G.~C. 1993, \aj, 105, 1495

\bibitem[{{Jackson} {et~al.}(1984){Jackson}, {Armstrong}, \&
  {Barrett}}]{1984ApJ...280..608J}
{Jackson}, J.~M., {Armstrong}, J.~T., \& {Barrett}, A.~H. 1984, \apj, 280, 608

\bibitem[{{Kraus} {et~al.}(2010){Kraus}, {Hofmann}, {Menten}, {Schertl},
  {Weigelt}, {Wyrowski}, {Meilland}, {Perraut}, {Petrov}, {Robbe-Dubois},
  {Schilke}, \& {Testi}}]{2010Natur.466..339K}
{Kraus}, S., {Hofmann}, K., {Menten}, K.~M., {et~al.} 2010, \nat, 466, 339

\bibitem[{{Krumholz} {et~al.}(2009){Krumholz}, {Klein}, {McKee}, {Offner}, \&
  {Cunningham}}]{2009Sci...323..754K}
{Krumholz}, M.~R., {Klein}, R.~I., {McKee}, C.~F., {Offner}, S.~S.~R., \&
  {Cunningham}, A.~J. 2009, Science, 323, 754

\bibitem[{{Kuiper} {et~al.}(2010){Kuiper}, {Klahr}, {Beuther}, \&
  {Henning}}]{2010ApJ...722.1556K}
{Kuiper}, R., {Klahr}, H., {Beuther}, H., \& {Henning}, T. 2010, \apj, 722,
  1556

\bibitem[{{Lee} {et~al.}(2005){Lee}, {Ho}, \& {White}}]{2005ApJ...619..948L}
{Lee}, C., {Ho}, P.~T.~P., \& {White}, S.~M. 2005, \apj, 619, 948

\bibitem[{{Leurini} {et~al.}(2009){Leurini}, {Codella}, {Zapata}, {Belloche},
  {Stanke}, {Wyrowski}, {Schilke}, {Menten}, \&
  {G{\"u}sten}}]{2009A&A...507.1443L}
{Leurini}, S., {Codella}, C., {Zapata}, L.~A., {et~al.} 2009, \aap, 507, 1443

\bibitem[{{Leurini} {et~al.}(2011){Leurini}, {Pillai}, {Stanke}, {Wyrowski},
  {Testi}, {Schuller}, {Menten}, \& {Thorwirth}}]{17233_irdc}
{Leurini}, S., {Pillai}, T., {Stanke}, T., {et~al.} 2011, subm. to \aap

\bibitem[{{M\" uller} {et~al.}(2005){M\" uller}, {Schl{\"o}der}, {Stutzki}, \&
  {Winnewisser}}]{2005JMoSt.742..215M}
{M\" uller}, H.~S.~P., {Schl{\"o}der}, F., {Stutzki}, J., \& {Winnewisser}, G.
  2005, Journal of Molecular Structure, 742, 215

\bibitem[{{M\" uller} {et~al.}(2001){M\" uller}, {Thorwirth}, {Roth}, \&
  {Winnewisser}}]{2001A&A...370L..49M}
{M\" uller}, H.~S.~P., {Thorwirth}, S., {Roth}, D.~A., \& {Winnewisser}, G.
  2001, \aap, 370, L49

\bibitem[{{MacLeod} {et~al.}(1998){MacLeod}, {Scalise}, {Saedt}, {Galt}, \&
  {Gaylard}}]{1998AJ....116.1897M}
{MacLeod}, G.~C., {Scalise}, E.~J., {Saedt}, S., {Galt}, J.~A., \& {Gaylard},
  M.~J. 1998, \aj, 116, 1897

\bibitem[{{Mart{\'{\i}}n} {et~al.}(2008){Mart{\'{\i}}n}, {Requena-Torres},
  {Mart{\'{\i}}n-Pintado}, \& {Mauersberger}}]{2008ApJ...678..245M}
{Mart{\'{\i}}n}, S., {Requena-Torres}, M.~A., {Mart{\'{\i}}n-Pintado}, J., \&
  {Mauersberger}, R. 2008, \apj, 678, 245

\bibitem[{{McKee} \& {Tan}(2002)}]{2002Natur.416...59M}
{McKee}, C.~F. \& {Tan}, J.~C. 2002, \nat, 416, 59

\bibitem[{{McKee} \& {Tan}(2003)}]{2003ApJ...585..850M}
{McKee}, C.~F. \& {Tan}, J.~C. 2003, \apj, 585, 850

\bibitem[{{Menten}(1991)}]{1991ApJ...380L..75M}
{Menten}, K.~M. 1991, \apjl, 380, L75

\bibitem[{{Miettinen} {et~al.}(2006){Miettinen}, {Harju}, {Haikala}, \&
  {Pomr{\'e}n}}]{2006A&A...460..721M}
{Miettinen}, O., {Harju}, J., {Haikala}, L.~K., \& {Pomr{\'e}n}, C. 2006, \aap,
  460, 721

\bibitem[{{Olmi} {et~al.}(1993){Olmi}, {Cesaroni}, \&
  {Walmsley}}]{1993A&A...276..489O}
{Olmi}, L., {Cesaroni}, R., \& {Walmsley}, C.~M. 1993, \aap, 276, 489

\bibitem[{{Pickett} {et~al.}(1998){Pickett}, {Poynter}, {Cohen}, {Delitsky},
  {Pearson}, \& {Muller}}]{pickett_JMolSpectrosc_60_883_1998}
{Pickett}, H.~M., {Poynter}, I.~R.~L., {Cohen}, E.~A., {et~al.} 1998, Journal
  of Quantitative Spectroscopy and Radiative Transfer, 60, 883

\bibitem[{{Rodr{\'{\i}}guez-Fern{\'a}ndez}
  {et~al.}(2010){Rodr{\'{\i}}guez-Fern{\'a}ndez}, {Tafalla}, {Gueth}, \&
  {Bachiller}}]{2010A&A...516A..98R}
{Rodr{\'{\i}}guez-Fern{\'a}ndez}, N.~J., {Tafalla}, M., {Gueth}, F., \&
  {Bachiller}, R. 2010, \aap, 516, A98

\bibitem[{{Walsh} {et~al.}(1998){Walsh}, {Burton}, {Hyland}, \&
  {Robinson}}]{1998MNRAS.301..640W}
{Walsh}, A.~J., {Burton}, M.~G., {Hyland}, A.~R., \& {Robinson}, G. 1998,
  \mnras, 301, 640

\bibitem[{{Wyrowski} {et~al.}(1999){Wyrowski}, {Schilke}, {Walmsley}, \&
  {Menten}}]{1999ApJ...514L..43W}
{Wyrowski}, F., {Schilke}, P., {Walmsley}, C.~M., \& {Menten}, K.~M. 1999,
  \apjl, 514, L43

\bibitem[{{Zapata} {et~al.}(2009){Zapata}, {Ho}, {Schilke}, {Rodr{\'{\i}}guez},
  {Menten}, {Palau}, \& {Garrod}}]{2009ApJ...698.1422Z}
{Zapata}, L.~A., {Ho}, P.~T.~P., {Schilke}, P., {et~al.} 2009, \apj, 698, 1422

\bibitem[{{Zapata} {et~al.}(2008){Zapata}, {Leurini}, {Menten}, {Schilke},
  {Rolffs}, \& {Hieret}}]{2008AJ....136.1455Z}
{Zapata}, L.~A., {Leurini}, S., {Menten}, K.~M., {et~al.} 2008, \aj, 136, 1455

\bibitem[{{Zapata} {et~al.}(2010{\natexlab{a}}){Zapata}, {Schmid-Burgk}, \&
  {Menten}}]{2010arXiv1009.1426Z}
{Zapata}, L.~A., {Schmid-Burgk}, J., \& {Menten}, K.~M. 2010{\natexlab{a}},
  ArXiv e-prints

\bibitem[{{Zapata} {et~al.}(2010{\natexlab{b}}){Zapata}, {Tang}, \&
  {Leurini}}]{2010ApJ...725.1091Z}
{Zapata}, L.~A., {Tang}, Y., \& {Leurini}, S. 2010{\natexlab{b}}, \apj, 725,
  1091

\bibitem[{{Zhang} {et~al.}(1998){Zhang}, {Hunter}, \&
  {Sridharan}}]{1998ApJ...505L.151Z}
{Zhang}, Q., {Hunter}, T.~R., \& {Sridharan}, T.~K. 1998, \apjl, 505, L151

\end{thebibliography}
\end{document}